\documentclass[final,5p,times,twocolumn]{elsarticle}
\usepackage{xspace,graphicx,color,pstricks,pst-plot,amssymb,amsmath,hyperref}
 
 \biboptions{sort&compress}
\newcommand{\jpsi}{\ensuremath{{\rm J}/\psi}\xspace}
\newcommand{\psip}{\ensuremath{\psi{\rm (2S)}}\xspace}
\newcommand{\chic}{\ensuremath{\chi_{c}}\xspace}
\newcommand{\chicZero}{\ensuremath{\chi_{c0}}\xspace}
\newcommand{\chicOne}{\ensuremath{\chi_{c1}}\xspace}
\newcommand{\chicTwo}{\ensuremath{\chi_{c2}}\xspace}
\newcommand{\upsAll}{\ensuremath{\Upsilon}\xspace}
\newcommand{\upsOne}{\ensuremath{\Upsilon{\rm (1S)}}\xspace}
\newcommand{\upsTwo}{\ensuremath{\Upsilon{\rm (2S)}}\xspace}
\newcommand{\upsThree}{\ensuremath{\Upsilon{\rm (3S)}}\xspace}
\newcommand{\upsAllS}{\ensuremath{\Upsilon{\rm (nS)}}\xspace}
\newcommand{\chib}{\ensuremath{\chi_{b}}\xspace}

\newcommand{\chiOne}{\ensuremath{\chi_{1}}\xspace}
\newcommand{\chiTwo}{\ensuremath{\chi_{2}}\xspace}
\newcommand{\oneSzero}{\ensuremath{{^1{\rm S}_0^{[8]}}}\xspace}
\newcommand{\threeSone}{\ensuremath{{^3{\rm S}_1^{[8]}}}\xspace}
\newcommand{\threePJ}{\ensuremath{{^3{\rm P}_J^{[8]}}}\xspace}
\newcommand{\threeSoneSinglet}{\ensuremath{{^3{\rm S}_1^{[1]}}}\xspace}

\newcommand{\pt}{\ensuremath{p_{\rm T}}\xspace}
\newcommand{\QQbar}{\ensuremath{Q \overline{Q}}\xspace}
\newcommand{\pTovM}{\ensuremath{p_{\rm T}/M}\xspace}
\newcommand{\chictwooverchicone}{\ensuremath{\chi_{c2}/\chi_{c1}}\xspace}
\newcommand{\chibtwooverchibone}{\ensuremath{\chi_{b2}/\chi_{b1}}\xspace}
\newcommand{\chitwooverchione}{\ensuremath{\chi_{2}/\chi_{1}}\xspace}
\newcommand{\chicoverjpsi}{\ensuremath{\chi_{c}/{\rm J}/\psi}\xspace}
\newcommand{\lth}{\ensuremath{\lambda_\vartheta}\xspace}

\journal{Physics Letters B}

\begin{document}
\begin{frontmatter}

\title{Quarkonium production at the LHC:\\
a phenomenological analysis of surprisingly simple data patterns}

\author[ist,lip]{Pietro Faccioli\corref{cor}}
\author[ist,cern]{Mariana Ara\'ujo}
\author[cern]{Valentin Kn\"unz}
\author[hephy]{Ilse Kr\"atschmer}
\author[cern]{Carlos Louren\c{c}o\corref{cor}} 
\author[ist,lip]{Jo\~ao Seixas}

\address[ist]{Physics Department, Instituto Superior T\'ecnico (IST), Lisbon, Portugal}
\address[lip]{Laborat\'orio de Instrumenta\c{c}\~ao e F\'{\i}sica Experimental de Part\'{\i}culas (LIP), Lisbon, Portugal}
\address[cern]{European Organization for Nuclear Research (CERN), Geneva, Switzerland}
\address[hephy]{Institute of High Energy Physics (HEPHY), Vienna, Austria}

\cortext[cor]{Corresponding authors: pietro.faccioli@cern.ch, carlos.lourenco@cern.ch}

\begin{abstract}
The LHC quarkonium production measurements reveal a startling observation: 
the J/$\psi$, $\psi$(2S), $\chi_{c1,2}$ and $\Upsilon$(nS) 
\pt-differential cross sections are compatible with one universal momentum scaling pattern. 
Considering also the absence of strong polarizations of directly and indirectly produced \mbox{S-wave} mesons,
we are led to the conclusion
that there is currently no evidence of a dependence of the partonic production mechanisms 
on the quantum numbers and mass of the final state.
The experimental observations supporting this universal production scenario are remarkably significant,
as shown by a new analysis approach, 
unbiased by specific theoretical calculations of partonic cross sections,
which are only considered a posteriori, in comparisons with the data-driven results.
\end{abstract}

\end{frontmatter}

\sloppy

\section{Introduction, motivation and conceptual remarks}
\label{sec:intro}

Most of the LHC analyses are devoted to ``searches"; 
hundreds of papers report studies aimed at verifying if the LHC data 
agree with hypotheses put forward by theory models. 
Such analyses give the primary role to the theory, 
which makes predictions based on calculations with different levels of accuracy and/or precision: 
some calculations are performed under assumptions that often incorporate approximations 
and only up to a given fixed order in a perturbative series.
In the case of heavy quarkonium production, 
non-relativistic quantum chromodynamics (NRQCD)~\cite{bib:NRQCD} is the commonly-considered standard theory, 
to be probed by experimental analyses. 
This model is, indeed, the most sophisticated, complex, and conceptually profound presently available 
in this chapter of physics. 
It is however also true that the level of NRQCD calculations is not yet sufficiently satisfactory, 
and new improvements are
continuously taking place.
The absence of a reliable ``standard model" is not, in itself, a problem, 
as long as we are aware of the transient nature of the calculations. 
Indeed, when experimental measurements are fitted (a suited word) within a given theoretical framework, 
we might be placing the data in a tight corset that moulds the patterns into desired shapes, 
preventing us from exploring a richer spectrum of options and, worse, 
potentially blinding us from the simplest and most natural interpretations.

In a previous publication~\cite{bib:FaccioliPLB736}, we showed how easy it is to create puzzles
when conducting global fits of quarkonium production data in the framework of NRQCD. 
All we need is to believe that a certain superposition of the presently available next-to-leading order (NLO) perturbative QCD calculations 
of short-distance coefficients (SDCs) must be able to describe the measured (unpolarized) differential cross sections 
down to very low quarkonium transverse momentum (\pt). 
The highest (statistical) precision of the lowest-\pt data drives the result of the fit 
and leads to an ``inescapable prediction": quarkonium production must be transversely polarized, already at not-so-high \pt values. 
The puzzling nature of the measured (absence of) quarkonium polarizations, which clearly contradict the predictions, 
can be trivially understood by simply realising that the existing NLO SDCs are not sufficiently accurate at low \pt. 
Indeed, Ref.~\cite{bib:FaccioliPLB736} shows that the existing SDCs are perfectly able to simultaneously describe, with a very good fit quality, 
the cross section and polarization data, provided we avoid the lowest \pt region. 
The puzzle was not in the data but rather in the belief that the current level of calculations already reached a sufficiently-high accuracy, 
even at low \pt.
If we assume that the theoretical calculations are accurate, we must conclude that the theory has been falsified by the LHC measurements 
(the corresponding global fits have astonishingly poor fit $\chi^2$ probabilities). 
Instead, we can reproduce all the existing data if we simply assume that the present calculations are inaccurate, 
a conclusion that keeps the door open regarding the validity of the theory at the fundamental level.

The study reported in the present Letter goes one step further in our data-driven approach: 
while respecting the NRQCD conceptual ideas, we reconsider its (hierarchical) boundaries
and allow ourselves to explore a broader landscape, 
guided by the map that Nature placed at the disposal of those who give her the central stage and who humbly try to understand her language. 
We are fortunate to have access to the treasure map uncovered by the LHC experiments, 
which provides clear and easy-to-follow indications to those who embrace an unbiased vision.
Indeed, the (high-\pt) quarkonium measurements provided by the LHC experiments 
reveal strong model-independent indications regarding the mechanisms of prompt quarkonium production, 
somehow missed until now,
presumably because the crystal-clear experimental patterns are obscured when seen through theoretically-driven perspectives.
These considerations have physical and methodological consequences. 
On one hand, such indications must provide inspiration for developments in the theoretical description of quarkonium production.
On the other hand, the quality reached by the experimental information suggests a new strategy for theory-data comparisons, 
where fits to measurements are performed with minimal theoretical ingredients 
and the results are compared only a posteriori with theory predictions based on fixed-order perturbative calculations.

In Section~\ref{sec:data} we present our central observation, 
brought forth by a comparative analysis of prompt quarkonium cross section and polarization measurements 
in proton-proton collisions at the LHC, with no recourse to model considerations: 
the data are compatible with a surprisingly simple scenario, 
in which production and decay kinematics follow an almost universal trend,
with no significant distinctions between states of different quantum numbers.
This observation raises the central physics question of this paper: how different are
the production mechanisms behind ${^3{\rm S}_1}$ and ${^3{\rm P}_J}$ quarkonia?
In Section~\ref{sec:theory} we discuss how the seemingly universal data patterns compare to the NRQCD theory framework,
with its relatively complex structure of hierarchies and constraints, 
and startling differences in the calculated kinematic behaviours of the participating processes.

We then discuss, in Section~\ref{sec:universalscenario}, a data-driven scenario.
While the production of ${^3{\rm S}_1}$ quarkonia can be described by a completely general parametrization, 
exploiting the detailed experimental information available on cross sections and polarizations, 
parametrizing the production of ${^3{\rm P}_J}$ states still requires model assumptions,
given the lack of corresponding polarization data.
Our scenario is a direct formalization of the measured trends, 
assuming universal production and decay properties for the ${^3{\rm S}_1}$ and ${^3{\rm P}_J}$ quarkonia. 
This represents an important change with respect to the hierarchy of elementary mechanisms foreseen by NRQCD, 
where $\chi$ production has, a priori, a very different process composition with respect to the
$\psi$ and $\Upsilon$ \mbox{S-wave} states.
We test this hypothesis in a global fit of charmonium data, 
properly accounting for the relevant feed-down decays. 
We report the general kinematic relations used for the modelling of the momentum distributions 
and polarizations of the indirectly produced states.
We also describe the details of our original data-fitting approach,
where the individual physical contributions to quarkonium production are exclusively discriminated by their characteristic polarizations, 
while the \pt distributions and relative normalizations are parametrized by a flexible, empirical function. 
Perturbative calculations of the production kinematics are not used as ingredients anywhere in our analysis. 
This study shows that it is possible to entrust the measurements, given their current level of precision, 
with the full responsibility of determining the physical outcome of the fit, obtaining reliable and significant results.
The theory calculations are only used a posteriori, in comparisons,
for each individual process, with the distributions determined by the fit.

\begin{figure*}[h!]
\centering
\includegraphics[width=0.85\textwidth]{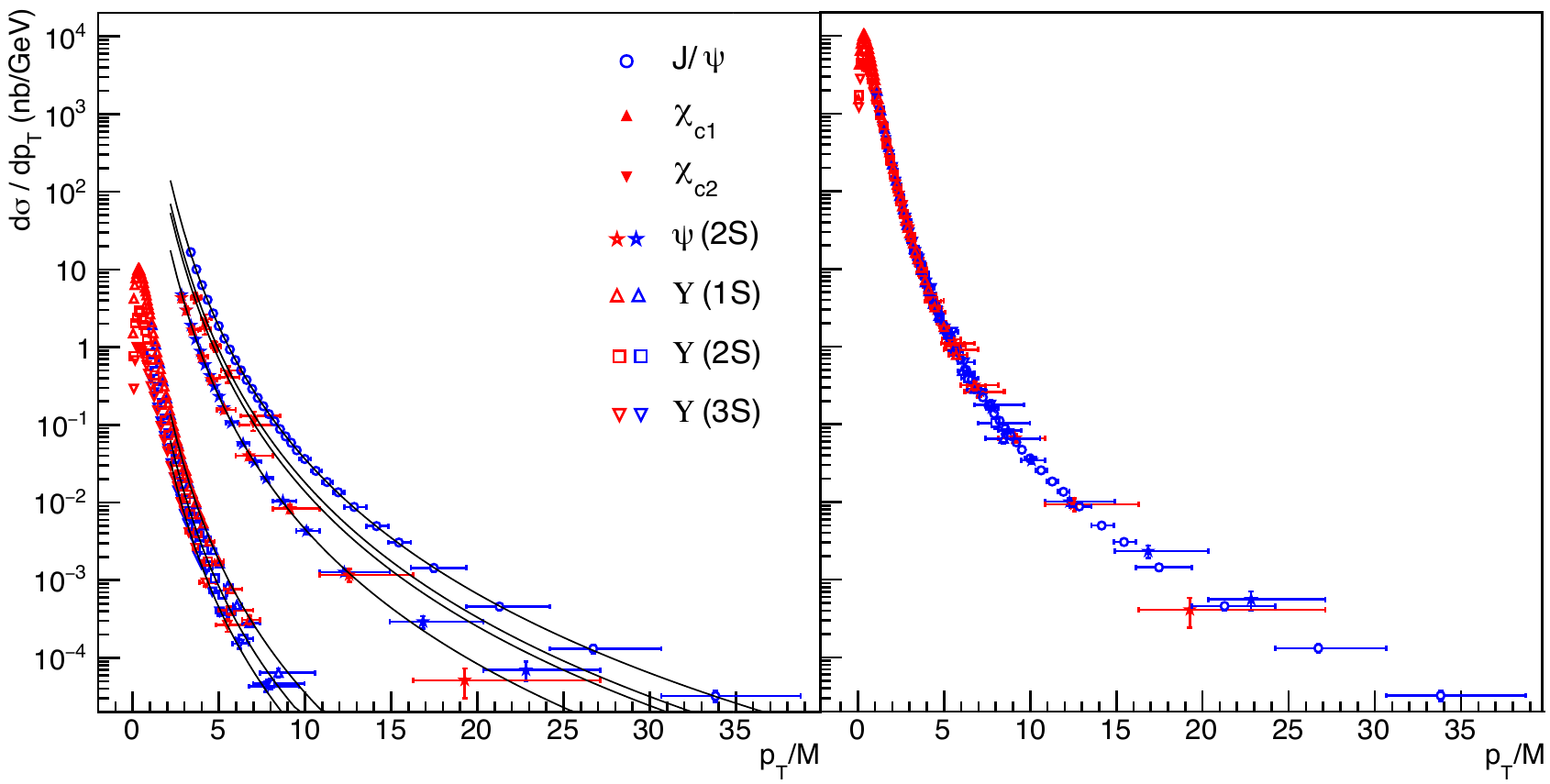}
\caption{Mid-rapidity prompt quarkonium cross sections measured in pp collisions 
at $\sqrt{s} = 7$\,TeV 
by ATLAS (red markers)~\cite{bib:ATLASpsi2S, bib:ATLASYnS, bib:ATLASchic}
and CMS (blue markers)~\cite{bib:CMSjpsi, bib:CMSYnS}, 
as a function of \pTovM.
The curves represent a single empirical function, 
with shape parameters determined by a simultaneous fit to all data (of $\pTovM > 3$)
and normalizations specific to each state (left panel) 
or adjusted to the \jpsi\ points (right panel) 
to directly illustrate the universality of the kinematic dependences.}
\label{fig:pTovM}
\end{figure*}

%%%%%%%%%%%%%%%%%%%%%%%%%%%%%%%%%%%%
\section{The surprising simplicity of the measured patterns}
\label{sec:data}

Figure~\ref{fig:pTovM} shows the first of the two interesting observations we want to discuss: 
a seemingly universal pattern in the shapes of the \pt\ distributions of all prompt quarkonia. 
Indeed, when presented as \pTovM\ distributions,
where $M$ is the mass of the quarkonium state, 
the prompt \jpsi, \chicOne, \chicTwo, \psip and \upsAllS production cross sections 
are all compatible with a single kinematic dependence, 
at least for mid-rapidity ($|y| \lesssim 1$) and for not-too-small \pTovM. 
This observation confirms, with higher-\pt data, a trend first noticed in Ref.~\cite{bib:FaccioliPLB736}. 
The consistency of the \pTovM\ distributions is quantified in Section~\ref{sec:universalscenario},
while Ref.~\cite{bib:Paper2} offers a detailed evaluation of 
how precisely the current cross section measurements constrain 
the \chic\ \pTovM\ trend to be similar to those of the \jpsi\ and \psip.

\begin{figure}[t]
\centering
\includegraphics[width=0.85\linewidth]{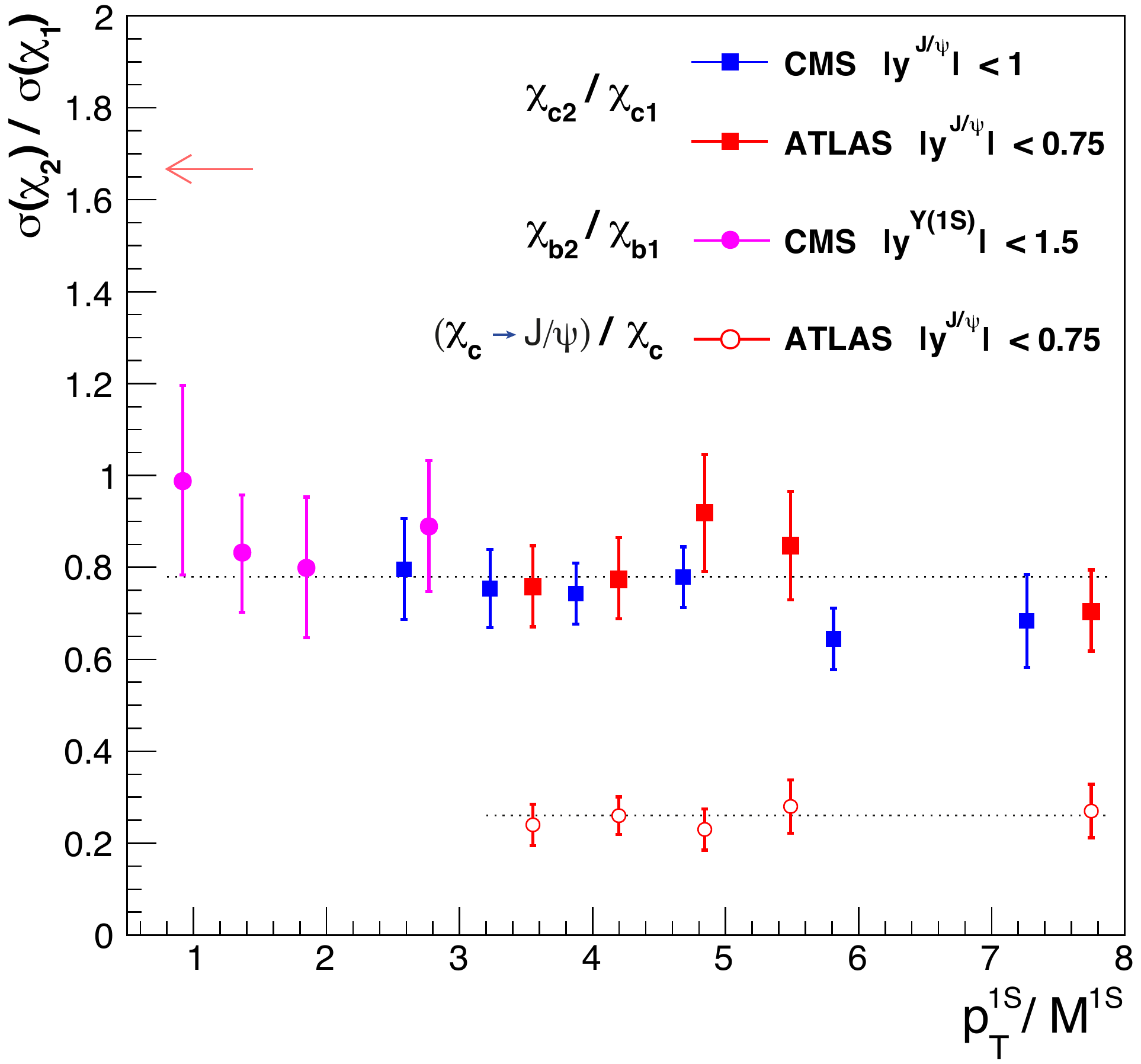}
\caption{Mid-rapidity \chictwooverchicone and \chibtwooverchibone
prompt yield ratios, as well as the fraction of the prompt \jpsi\ yield coming from \chic\ decays,
measured in pp collisions by ATLAS~\cite{bib:ATLASchic} 
and CMS~\cite{bib:CMSchic, bib:CMSchib}, 
as a function of \pTovM. 
The arrow indicates $5/3$, 
the value predicted by heavy-quark spin-symmetry for the \chictwooverchicone and \chibtwooverchibone ratios 
in the hypothesis of pure colour-octet production.}
\label{fig:chiratios}
\end{figure}

Figure~\ref{fig:chiratios} presents a clearer view of the \chic\ measurements, also adding \chib\ data: 
the \chictwooverchicone\ and \chibtwooverchibone\ yield ratios, 
as well as the ratio of \jpsi\ from \chic\ decays to prompt \jpsi\ yield, 
show a flat dependence vs.\ \pTovM. 
Since the prompt \psip mesons are fully directly produced 
while the \jpsi\ and \upsAllS\ states are significantly affected by feed-down contributions from \chic\ and \chib\ decays
($\simeq 25\%$~\cite{bib:ATLASchic} and $\simeq 40\%$~\cite{bib:LHCbChibFeedown}, respectively),
the perfect compatibility of their \pTovM\ shapes is another observation supporting the similarity 
of P- and \mbox{S-wave} quarkonium production, even in \pTovM\ ranges uncovered by the existing $\chi$ data.

The simplest explanation of such universality of the production kinematics is that 
one and the same parton-level process (or mixture of processes) 
describes the production of all states, irrespectively of their masses and quantum numbers. 
In fact, the kinematic dependence of the partonic cross section of a given process is invariant 
by simultaneous rescaling of all energy-related variables. 
This translates to an invariance by \pTovM\ rescaling, 
if the longitudinal momentum components can be neglected:
such scaling is not expected to be valid in the low \pTovM\ and high-rapidity
region of the LHCb 
%data~\cite{bib:LHCb_psi_cs, bib:LHCb_psip_cs, 
%bib:LHCb_psi_pol, bib:LHCb_psip_pol, bib:LHCb_chicRatio, bib:LHCb_chicJpsiRatio}.
data (see Refs.~\cite{bib:LHCb_chicRatio, bib:LHCb_chicJpsiRatio, bib:LHCb_psip_pol} and references therein).
On the other hand, as shown in Ref.~\cite{bib:FaccioliPLB736}, 
current fixed-order perturbative calculations are seemingly unsuitable 
to describe cross section measurements for $\pTovM \lesssim 3$, 
justifying that relatively high-\pt\ data remain preferable in data-theory comparisons.

\begin{figure}[t]
\centering
\includegraphics[width=0.85\linewidth]{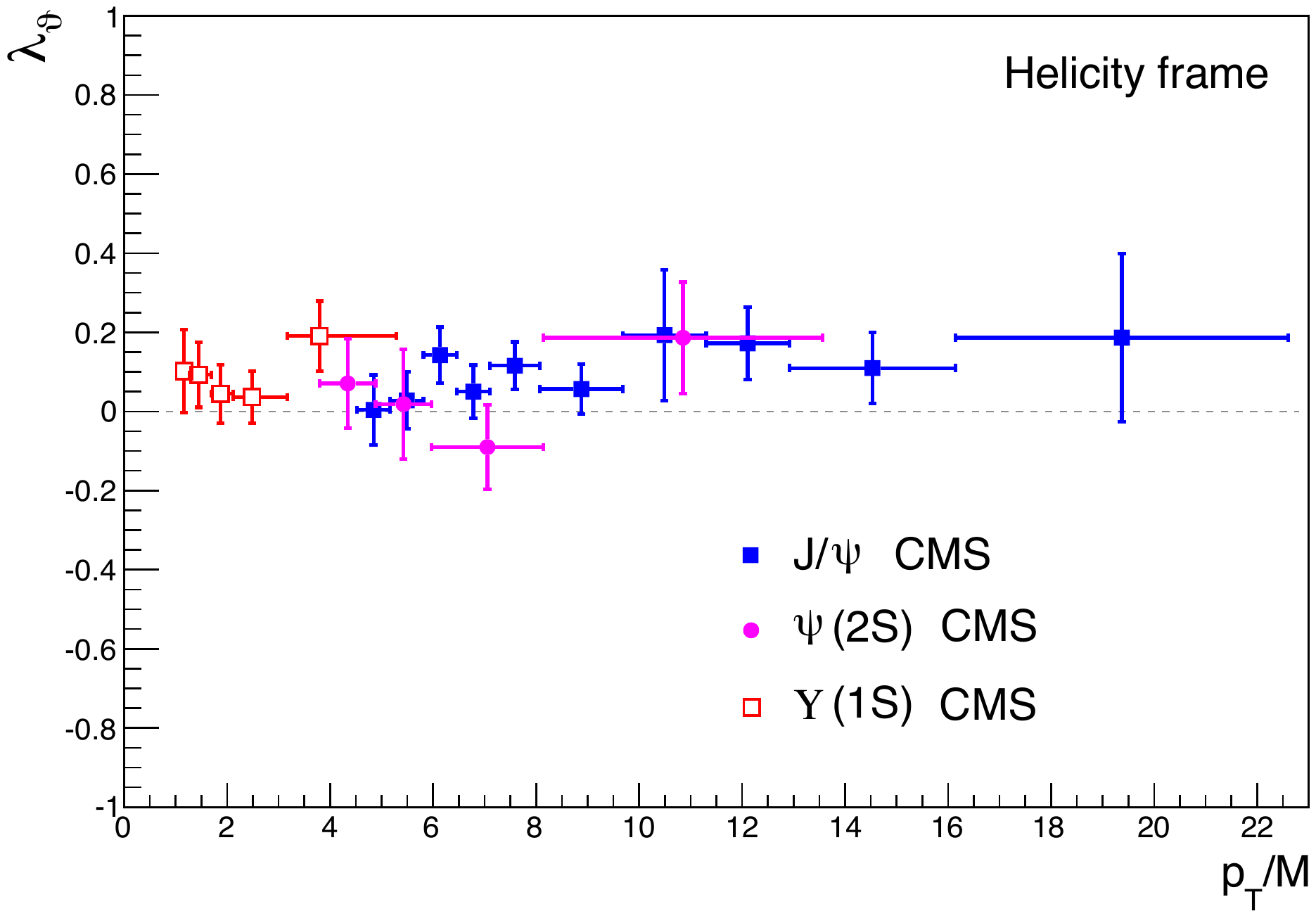}
\caption{Polar anisotropy parameter $\lambda_\vartheta$ 
measured by CMS in pp collisions at $\sqrt{s} = 7$\,TeV,
for prompt \jpsi, \psip and \upsOne dilepton decays~\cite{bib:CMSlambdaPsi2S,bib:CMSlambdaYnS}. 
For improved clarity, values corresponding to two or three rapidity bins were averaged, 
assuming uncorrelated systematic uncertainties, 
and the very uncertain \upsTwo and \upsThree data are not shown.}
\label{fig:polarizations}
\end{figure}

The second experimental result guiding our analysis is the absence of significant polarizations 
in the measured quarkonium decay distributions, often considered a puzzling observation given that
most theory predictions point to significant polarizations, increasing with \pt~\cite{bib:EPJC69}.
The most precise measurements of the $\lambda_\vartheta$ polar anisotropy parameter
of the \jpsi, \psip\ and \upsAllS\ dilepton decay distributions, 
shown in Fig.~\ref{fig:polarizations} (for the helicity frame) as a function of \pTovM\ and averaged over rapidity, 
are compatible with a small (even negligible) and constant polarization. 
While no direct $\chi$ polarization measurements exist, 
we remind that the three states shown in Fig.~\ref{fig:polarizations} 
have very different $\chi$ feed-down components, 
suggesting that weak polarizations are also to be expected for the \chic\ and \chib\ states.

We conclude that, today, there is no experimental evidence of differences in production and decay kinematics 
between quarkonium states of different masses and angular momentum properties, 
at least in the mid-rapidity region. 
Such a ``universal production'' scenario is a surprising experimental observation: 
in principle, conservation rules should make partonic production cross sections
different for states of different quantum numbers. 
In the next section we discuss how the current models of quarkonium production relate to this observation.

%%%%%%%%%%%%%%%%%%%%%%%%%%%%%%%%%%%%
\section{The not-so-simple theory patterns}
\label{sec:theory}

Current studies of quarkonium production phenomenology are based on NRQCD~\cite{bib:NRQCD}, 
a non-relativistic effective field theory whose pillar is the 
hypothesis of factorization of the long- and short-distance parts of the production process. 
Under this hypothesis, the inclusive prompt production cross section of a quarkonium state $H$, 
after a collision of initial systems $A$ and $B$, can be written as a linear combination 
of SDCs ($\mathcal{S}$) to produce heavy quark-antiquark pairs (\QQbar) 
of different colours (singlet or octet) and spin-angular momentum configurations:
\begin{align}
\label{eq:factorization}
\begin{split}
  \sigma(A+B \to H+X) = 
  \sum\limits_{S,L,C}
  & \mathcal{S}(A+B\to \QQbar[^{2S+1}L_J^C] +X) \\
  & \times \mathcal{L}( \QQbar[^{2S+1}L_J^C] \to H)  \; .
\end{split}
\end{align}

The coefficients $\mathcal{L}$ are the so-called long-distance matrix elements (LDMEs), 
representing the probabilities that the different \QQbar\ states, 
with spin, angular-momentum and colour configuration described by the indices $S$, $L$, $J$ and $C$, 
evolve into the observable quarkonium. 
The LDMEs are assumed to be universal constants, for a given quarkonium,
independent of the production mechanism and kinematics (collision system and energy, \pt and rapidity),
while the SDCs are functions of the collision type and energy, as well as of the \QQbar\ laboratory momentum, 
and are calculated perturbatively.
The LDME coefficients can be determined by comparison with the quarkonium cross sections measured as a function of \pt. 
It is important to remember, however, that the LDME and the SDC of a given term in the expansion are, individually, 
unobservable, as their separation depends on the NRQCD factorization scale. 
Also the intermediate \QQbar\ pre-resonance state should not be considered a physical state. 
The fact that the LDMEs can be determined using experimental data does not mean that they are physical observables: 
their extraction presumes the knowledge of (equally unphysical) SDC functions. 
This also means that the LDME values, as determined from data, 
depend on the perturbative order of the SDC calculations and, effectively, lose their universal character, 
given that the evolution of perturbative calculations through successive orders 
usually follows different patterns for different collision systems and/or \pt\ and rapidity regions.

Assuming the limit of heavy component quarks strongly reduces and constrains,
for each quarkonium,
the wide and indiscriminate range of processes entering Eq.~\ref{eq:factorization}.
In fact, for small values of the relative velocity, $v$, of the two heavy quarks (non-relativistic limit), 
a strong hierarchy arises in the LDME magnitudes and only a small number of dominating terms, 
depending on the final state, are kept in the linear combination.
Moreover, some of these terms are ruled out by the perturbative calculations of the corresponding SDCs, 
because they lead to negligible yields in comparison to the observed prompt cross sections, 
a well known case being the \threeSoneSinglet\ 
singlet contribution to \jpsi\ and \psip\ production~\cite{bib:CDFpsisRunI}.
As a result, the prompt production cross sections of the ${^3{\rm S}_1}$ quarkonia, \jpsi, \psip\ and \upsAllS,
are expected to be dominated by the \oneSzero, \threeSone and \threePJ octet contributions, 
with comparable magnitudes (the \threeSone\ being suppressed at the short-distance level), 
the \chic\ and \chib\ production should mainly proceed through the \threeSone and ${^3{\rm P}_{1,2}^{[1]}}$ channels, 
and the $\eta_c$ production should reflect the ${^1{\rm S}_{0}^{[1]}}$ singlet and the \threeSone octet channels 
(the \oneSzero and ${^1{\rm P}_{1}^{[8]}}$ terms being suppressed by the small SDC times LDME 
products~\cite{bib:etac_Kniehl}).

Besides \mbox{$v$-scaling}, 
heavy-quark spin-symmetry (HQSS) also plays a role in constraining the relative magnitudes of the NRQCD LDMEs. 
For example, the ratio between the octet LDMEs for $\chi_{c2,b2}$ and $\chi_{c1,b1}$ production 
is expected to be equal to the corresponding ratio of spin states,
\begin{equation}
\label{eq:FiveThirds}
\mathcal{L}(\threeSone \to \chi_{c2,b2}) \,/\, 
\mathcal{L}(\threeSone \to \chi_{c1,b1}) = 5/3 \; .
\end{equation}

\begin{figure*}[t]
\centering
\includegraphics[width=0.85\linewidth]{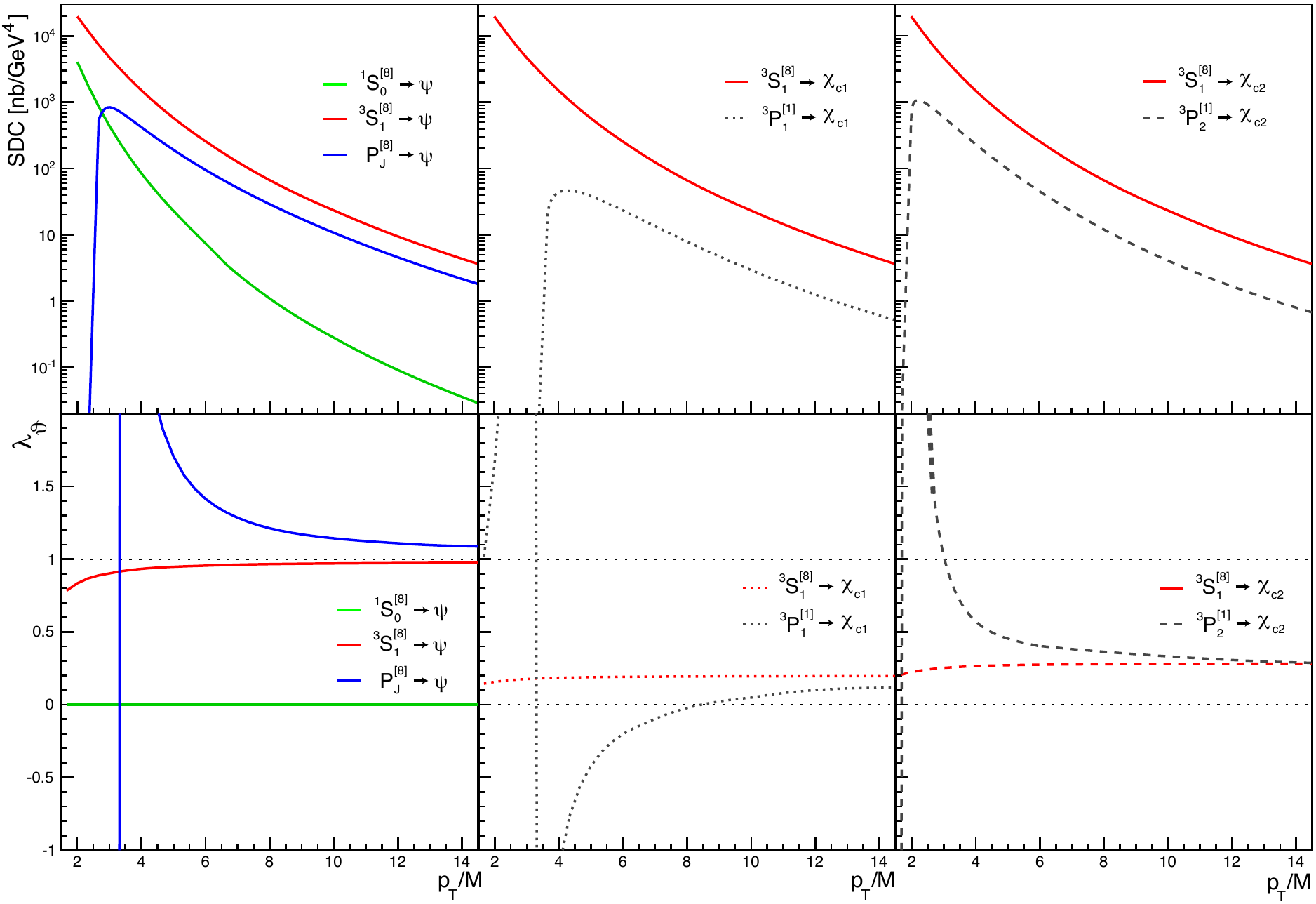}
\caption{SDCs (top) and corresponding helicity-frame $\lambda_\vartheta$ polarization parameters (bottom),
for the main octet and singlet components of directly-produced $\psi$ mesons (left), 
and for \jpsi mesons from $\chi_{c1}$ (middle) and $\chi_{c2}$ (right) decays,
calculated at NLO~\cite{bib:Chao:2012iv, bib:Shao:2014fca, bib:Shao:2015vga} 
for pp collisions at $\sqrt{s} = 7$\,TeV and a charmonium mass of 3\,GeV. 
The \threePJ SDC is defined as 
$\mathcal{S}({^3{\rm P}_{0}^{[8]}})+
3\mathcal{S}({^3{\rm P}_{1}^{[8]}})+
5\mathcal{S}({^3{\rm P}_{2}^{[8]}})$.
The \threePJ and ${^3{\rm P}_{1,2}^{[1]}}$ SDC functions are multiplied by $m_c^2$, the squared charm-quark mass;
they become negative at high \pTovM\ and are plotted with flipped signs.}
\label{fig:SDCs}
\end{figure*}

In the present study,
we are mainly interested in the characteristic kinematic dependencies of the different contributing processes 
and on how they compare to the simplicity of the experimental scenario. 
Figure~\ref{fig:SDCs} shows the individual contributions of the dominating \QQbar\ configurations, 
calculated at NLO~\cite{bib:Chao:2012iv, bib:Shao:2014fca, bib:Shao:2015vga}.
The top panels show the \pTovM\ dependence of each SDC;
the \mbox{P-wave} SDCs are shown (here and throughout the paper) multiplied by $m_c^2$.
The bottom panels show the polarization parameter $\lambda_\vartheta = (1- 3\, \beta_0)/(1+ \beta_0)$ 
for \mbox{S-wave} dilepton decays,
where $\beta_0$ is the longitudinal cross section fraction 
(angular momentum projection $J_z = 0$) in the helicity frame,
for directly-produced \jpsi or \psip mesons (left) 
and for mesons resulting from $\chi_{c1}$ (middle) or $\chi_{c2}$ (right) feed-down decays.

The \mbox{P-wave} contributions have rather peculiar kinematic behaviours,
with cross sections becoming negative above certain threshold \pTovM\ values 
and unphysical polarization parameters ($|\lambda_\vartheta|>1$). 
In fact, the calculations of the singlet and octet \mbox{P-wave} SDCs present singularities that match 
(and can be cancelled by) the one appearing in the \threeSone\ LDME~\cite{bib:NRQCD, bib:Petrelli}. 
Therefore, only the ${^3{\rm P}_J^{[1,8]}} + \threeSone$ sum is required to be well-defined and physical, 
while the two ``fragments'' ${^3{\rm P}_J^{[1,8]}}$ and \threeSone are, effectively, 
an internal mathematical detail of the calculation, their relative contribution depending, 
for example, on the choice of the unobservable NRQCD factorization scale. 
In other words, the individual terms of the calculation are, on their own, unobservable. 
In particular, the ${^3{\rm P}_J^{[1,8]}}$ components at NLO contain large and negative transverse and/or longitudinal contributions, 
leading to the kinematic peculiarities mentioned above. 
While such characteristics can be acknowledged as technical details of a theoretical calculation 
involving an expansion over purely mathematical objects, they are, nevertheless, 
in striking contradiction with the universality of the \pTovM\ dependencies 
of the cross sections and polarizations measured for the different quarkonium states. 
The non-trivial \pTovM\ dependencies and sign-changing normalizations of the \mbox{P-wave} contributions 
require seemingly miraculous cancellations to yield results that, 
besides being physical, must reproduce the comparatively trivial experimental trends.

The different final states come from different pre-resonance mixtures, 
of characteristic kinematic trends: 
the \threeSone octet term, containing the contributions of processes where the \QQbar\ comes from a single gluon,
is (at high-enough \pt) fully transversely polarized,
while the \oneSzero\ octet is unpolarized (by definition) and has a much steeper \pTovM\ shape.
At high \pTovM, all three \mbox{P-wave} components seem to asymptotically tend
to the cross section and polarization shapes of their complementary-term \threeSone.
At lower \pTovM, instead, in particular in the region covered by current $\chi$ measurements, 
the trends strongly deviate and, for example, 
the ${^3{\rm P}_1^{[1]}}$ and ${^3{\rm P}_2^{[1]}}$ terms sharply differentiate the $J = 1$ and 2 states,
possibly leading to strong polarizations. 
Comparing the SDCs of Fig.~\ref{fig:SDCs} with the experimental trends of Fig.~\ref{fig:chiratios}, 
we see that, in particular, the different turnover points of the $^3P_1^{[1]}$, $^3P_2^{[1]}$ and \threePJ SDCs 
may prevent a smooth description of the flat \chitwooverchione\ and \chicoverjpsi\ ratios. 
It is true that the relative importance of the \mbox{P-wave} contributions with respect to the other participating processes 
is a priori unknown and has to be tuned using the data. 
In NRQCD, however, \mbox{$v$-scaling} rules predict an approximate equality in magnitude 
between the \oneSzero, \threeSone and \threePJ LDMEs for the production of ${^3{\rm S}_1}$ quarkonia, 
and between the ${^3{\rm P}_{1,2}^{[1]}}$ and \threeSone ones for the production of $\chi$ mesons. 
Moreover, the HQSS relation in Eq.~\ref{eq:FiveThirds} would lead to
a \chitwooverchione\ ratio close to $5/3$ if the \threeSone octet term dominated $\chi$ production. 
The measured ratio (Fig.~\ref{fig:chiratios}) violates the octet-only expectation by a factor of 2, 
indicating that the singlet components must be very important. 
On the other hand, large singlet terms would lead to a significant difference between the $J = 1$ and 2
\pTovM dependencies, probably excluded by the measured ratio.

This picture gives the impression 
that the variety of pre-resonance contributions and their associated production and decay kinematics, 
as implied by \mbox{$v$-scaling} hierarchies, HQSS and current perturbative calculations, 
is redundant with respect to the observed universal \pTovM\ scaling and lack of polarization. 
The simple data patterns cannot be reproduced without invoking conspiring cancellations.
In other words, the current state-of-the-art theory does not seem to naturally accommodate the experimental scenario, 
whose simplicity would be interpreted, in this framework, as a coincidence. 
On the other hand, the responsibility of achieving the necessary cancellations is left to the LDMEs (Eq.~\ref{eq:factorization}), 
which are not calculated and must be obtained from global fits to measurements. 
This procedure may lead to unstable results, 
given that the fits are driven by the necessity of cancelling kinematic SDC dependencies not seen in the data. 
Seemingly successful cancellations may be the result of a temporary coincidence, allowed by the experimental uncertainties
on the observables most sensitive to the critical \mbox{P-wave} contributions:
the $\chi$ cross sections are still measured with a rather low precision with respect to the \jpsi\ and \psip\ cross sections, 
and no $\chi$ polarization measurements exist so far. 
The resulting LDMEs may, therefore, be artificial and unphysical,
a situation that cannot be immediately identified as a problem, 
given the absence of expected/calculated counterparts.

Furthermore, it is difficult to estimate the precision of the current SDC calculations, at least in the case of the \mbox{P-wave} terms. 
Such components show dramatic modifications from leading-order (LO) to NLO calculations,
changing, in particular, from positive SDCs and reasonable polarizations at LO to mostly negative SDCs and unphysical
polarizations at NLO~\cite{bib:Chao:2012iv, bib:Shao:2014fca, bib:Shao:2015vga, bib:BKNPB, bib:BKMPLA, bib:BKmodel}.
Moreover, leading-power fragmentation corrections~\cite{bib:BodwinCorrections},
accounting for a subset of next-to-next-to-leading-order (NNLO) processes, 
still bring startling shape and normalization modifications to the \threePJ SDC 
(no corresponding calculations exist for the ${^3{\rm P}_{1,2}^{[1]}}$ terms), 
indicating a slow convergence of the perturbative expansion.

It is important to emphasise the potential fragility of the fit procedures currently adopted in NRQCD analyses of quarkonium production. 
These analyses rely on free LDMEs, representing unphysical quantities not comparable to direct observations or calculations, 
to obtain precise cancellations between the individually-unobservable SDCs
of a mathematically-expanded physical cross section. 
Such cancellations must be accomplished over the entire kinematic phase space, 
while the LDMEs are independent of kinematics. 
The reliability of such procedures would require, in general, a highly over-constrained experimental scenario, 
not yet reached at present, especially in the \chic\ and \chib\ measurements and, to some extent, 
in all polarization measurements, which still lack precision and completeness. 
However, it is especially the precision of the SDC calculations that should be placed under control. 
Currently, no meaningful theoretical uncertainties are associated to the kinematic dependencies of the calculated SDCs, 
some of which show drastic shape modifications across perturbative orders. 
This is particularly worrisome given that significant deviations from ideal calculations may affect the results 
of the necessary term-to-term cancellations and even invalidate, in practice if not conceptually, 
the pivotal postulate that the coefficients of the factorization expansion are universal constants.

Motivated by these considerations, we adopt a completely data-driven analysis method,
to ensure that the fit outcome is stable and susceptible to definite physical interpretations, 
even if at the cost of larger uncertainties in the results and predictions, truthfully reflecting the current experimental inputs.
In this new fitting framework, only observable parameters and cross section contributions are considered,
the perturbative kinematic calculations being replaced by empirical functions, to be determined by the data. 

%%%%%%%%%%%%%%%%%%%%%%%%%%%%%%%%%%%%
\section{The universal almost unpolarized production scenario}
\label{sec:universalscenario}

Inspired by the scenario of maximum simplicity depicted by the LHC data (Section~\ref{sec:data}), 
in this section we discuss and test the correspondingly simple physical hypothesis 
that one common unpolarized mechanism dominates the production of both ${^3{\rm S}_1}$ and ${^3{\rm P}_J}$ quarkonium states,
while allowing the existence of a ``corrective'', transversely polarized term.
This ``universal almost unpolarized'' (UAU) production mechanism
does not obey the heavy-quark-limit hierarchies and constraints of NRQCD, 
in particular for what concerns the $\chi$ states.
In the framework of NRQCD, universal production naturally translates into the existence of a common colour-octet channel leading, 
with suitable changes of quantum numbers, to any final quarkonium state. 
With \chiOne\ and \chiTwo\ produced from the same octet pre-resonance state, 
HQSS would predict
a \chitwooverchione\ ratio of $5/3$, very different from the measured values,
which cluster around 0.8 (Fig.~\ref{fig:chiratios}). 
On the other hand,
it can be argued that this constraint on the ratio of octet components is only an approximation,
since it neglects effects related to the $\chiTwo-\chiOne$ mass difference and spin-orbital interactions,
possibly leading to unequal wave functions of \chiOne and \chiTwo~\cite{bib:BaranovChi2011,bib:BaranovChi2015}.
In fact, Eq.~\ref{eq:FiveThirds} has observable counterparts, 
for example, in the following ratios of branching fractions~\cite{bib:PDG}:
\begin{align}
\label{eq:FiveThirdsViolations}
\begin{split}
\psip \to \chi_{c2} \,\gamma \;/\;
\psip \to \chi_{c1} \,\gamma \quad~~~
&: 0.95 \pm 0.05 \; , \\ 
\upsTwo \to \chi_{b2}{\rm (1P)} \,\gamma \;/\; 
\upsTwo \to \chi_{b1}{\rm (1P)} \,\gamma
&: 1.04 \pm 0.08 \; , \\ 
\upsThree \to \chi_{b2}{\rm (2P)} \,\gamma \;/\;
\upsThree \to \chi_{b1}{\rm (2P)} \,\gamma 
&: 1.04 \pm 0.16 \; . 
\end{split}
\end{align}

All these ratios are significantly different from $5/3$ 
and, interestingly, much closer to the measured \chitwooverchione\ ratios shown in Fig.~\ref{fig:chiratios}.
Incidentally, the ratio 
$B(\psip \to \chi_{c2} \gamma) \,/\, B(\psip \to \chi_{c0} \gamma) = 1.10 \pm 0.05$~\cite{bib:PDG}
violates even more dramatically the corresponding spin-counting expectation of $5$.
Furthermore, another HQSS relation,  
$\mathcal{L}(\threeSone \to \eta_c) = \mathcal{L}(\oneSzero \to \jpsi)$, 
translates into the prediction of a large \threeSone contribution to $\eta_c$ production in pp collisions. 
But the pure ${^1{\rm S}_{0}^{[1]}}$ singlet cross section, a parameter-free NRQCD prediction, 
already adequately describes~\cite{bib:etac_Kniehl}, alone, 
the measured $\eta_c$ cross section~\cite{bib:LHCb_etac}.
The addition of the \threeSone component significantly overshoots the data, 
a ``puzzle'' created by the HQSS constraint. 
We also note that spin-counting rules have been seen to fail even in the description of the production of simpler, 
heavy-light quark systems, 
predicting for example wrong yield ratios between charged and neutral $D$ mesons~\cite{bib:PhysRepHF,bib:pv}.

The violation of HQSS-limit constraints is not the only source of conflict between the UAU hypothesis and NRQCD.
The natural candidate for a single mechanism capable of producing any final state leading to unpolarized production is the \oneSzero octet channel. 
In NRQCD, the role of this process in \mbox{P-wave} quarkonium production is disfavoured by the \mbox{$v$-scaling} rules, 
which favour the dominance of the combinations ${^3{\rm S}_1^{[8]}} + {^3{\rm P}_{1,2}^{[1]}}$,
characterized by non-zero polarizations and much flatter high-\pTovM\ distributions (Fig.~\ref{fig:SDCs}). 
In fact, a priori, 
the current SDC calculations predict a significant difference between the $\chi$ and $\psi$ production kinematics.
Incidentally, the dominance of \threeSone as common production channel would fit much better the \mbox{$v$-scaling} hierarchies, 
such process being already a leading contribution for both $\chi$ and $\psi$ production,
but this hypothesis leads to strong transverse polarizations of the ${^3{\rm S}_1}$ quarkonia and is excluded by the data.
In summary, when seen as a ``\oneSzero dominance model", 
the data-inspired UAU scenario represents a stretching of the 
heavy-quark-limit hierarchies of NRQCD towards an unforeseen, stronger hierarchy. 

While the UAU hypothesis may seem a rather specific physical assumption,
its implementation in our analysis is general, as much as allowed by the existing data. 
Inspired by the pattern of Fig.~\ref{fig:polarizations}, we consider that
the production cross section can be decomposed 
into an ``unpolarized'' term and a ``transversely polarized'' term.
These denominations are the only qualifications we use in our description of the participating processes. 
The two terms can be ideally associated, respectively, 
to the NRQCD \oneSzero octet channel and to the net contribution of all other mechanisms, 
but any other scenario would fit in this framework, 
including one involving a superposition of (unobserved) longitudinally and transversely polarized sub-processes, 
provided that the observable net result is not longitudinal. 
This procedure could be generalized to scenarios including observed longitudinal polarizations by replacing the unpolarized contribution with a longitudinally polarized one and letting the data determine, in a model-independent way, the longitudinal and transverse cross-sections contributions. 

In a production scenario characterized by non-longitudinal observable polarization,
the cross section for the \emph{direct} yield of a given ${^3{\rm S}_1}$ state can be parametrized as
\begin{equation}
\label{eq:DirectXsect}
\sigma_{\rm dir}(\pTovM) = \sigma_{\rm dir}^{*} [ (1-f_{\rm p}^{*}) \, g_{\rm u}(\pTovM) + f_{\rm p}^{*} \, g_{\rm p}(\pTovM) ] \;,
\end{equation}
where $\sigma_{\rm dir}^{*}$ and $f_{\rm p}^{*}$ are, respectively, 
the total direct-production cross section and the fractional contribution of the polarized process, 
both calculated at a fixed reference value, which we choose to be $(p_{\rm T}/M)^{*} = 2$. 
The \pTovM\ dependencies of the unpolarized and polarized yields are described by the shape functions 
$g_{\rm u}(\pTovM)$ and $g_{\rm p}(\pTovM)$, respectively,
both normalized to unity at $(p_{\rm T}/M)^{*}$,
$g(\pTovM) = h(\pTovM) / h((p_{\rm T}/M)^{*})$, with 
\begin{equation}
\label{eq:powerLaw}
h(\pTovM) =
\frac{p_{\rm T}}{M} \cdot \bigg( 1+\frac{1}{\beta-2} \cdot \frac{(p_{\rm T}/M)^2}{\gamma} \bigg)^{-\beta}\, .
\end{equation}

The parameter $\gamma$ defines the function  in the low-\pt\ turn-on region and is 
only mildly sensitive to the data we are considering here; 
hence, in the fit we consider $\gamma$ as a common free parameter.
The $\beta$ power-law exponent, instead, characterizes the high-\pt shape
($h \propto (p_{\rm T}/M)^{1-2\beta}$ for $\pTovM \gg \sqrt{\gamma (\beta-2)}$);
therefore, we 
distinguish the unpolarized and polarized cross sections with two different powers, 
$\beta_{\rm u}$ and $\beta_{\rm p}$, respectively.

The discrimination between the two physical contributions relies on the experimental polarization constraint, 
the unpolarized and polarized processes for ${^3{\rm S}_1}$ states being characterized  
by $\lambda_\vartheta = 0$ and $\lambda_\vartheta = 1$, respectively. 
The polarized yield fraction can be expressed, as a function of \pTovM, 
as $f_{\rm p} = 3 \lambda_\vartheta  / (4 - \lambda_\vartheta)$. 
This is a crucial difference of approach with respect to fits using SDC shapes fixed by calculations.
In those analyses, the result of the fit is mainly or exclusively determined by the comparison of the calculated \pt\ dependencies 
with the cross section measurements, while the less precise polarization data
are not included in the fits or have a relatively weak constraining effect. 
We adopt the opposite point of view, leaving to the polarization measurements, as functions of \pTovM, 
the exclusive role of constraining both the relative normalizations 
and the relative differences between the momentum dependencies of the different process contributions.
Given that the analysis becomes, therefore, fully driven by the experimental measurements, 
the precision of the results is no longer limited by unquantifiable theoretical uncertainties 
and will improve as the experimental panorama evolves.

Our previous analysis~\cite{bib:FaccioliPLB736}, using calculated SDCs, 
addressed the production of the \psip and \upsThree quarkonia, 
the ones less affected by feed-down decays from \mbox{P-wave} quarkonia. 
We now concentrate on the charmonium system, complementing the \psip measurements with \jpsi, \chicOne and \chicTwo data. 
The bottomonium system, with its more articulated feed-down structure, not yet sufficiently constrained by data 
(no mid-rapidity \chib\ measurements exist), is left for future consideration. 
We also profit from much improved \psip and \jpsi cross section measurements recently made available by ATLAS and CMS,
extending to significantly higher \pt. 
The similarity of the \pTovM\ dependencies shown by these precise measurements, 
discussed in Section~\ref{sec:data}, 
guides us to assuming that the direct production mechanism is the same for the \jpsi\ and \psip mesons. 
This translates, in our fit, in the use of a single ``polarized fraction'', represented by the parameter $f_{\rm p}^{*}$, common to the two states.

No significant experimental indications exist, so far, regarding \chic\ polarization. 
This observable may be indirectly constrained by the difference between the measured \jpsi\ and \psip polarizations, 
the former including about 25\% of \chic\ contribution while the latter is fully directly produced.
Future measurements of the \chic\ polarization and/or more precise ones of the \jpsi\ and \psip polarizations 
will allow us to perform a fully data-driven fit, 
without conjectures on differences between the \chic\ production and the \jpsi\ or \psip\ production, 
that is, without constraints on the ``universality'' of S- and \mbox{P-wave} quarkonium production.
At the present moment, however, the fit results are insensitive to different \chic\ polarization hypotheses.
To ensure that the fit is sufficiently constrained, given the precision of the existing data, 
we reduce the number of free parameters by adopting plausible relations between the properties of S- and \mbox{P-wave} states.
It is only at this point, and only for the production of \mbox{P-wave} quarkonia, 
that our fit framework acquires model-dependent ingredients.
While other hypotheses are considered in Ref.~\cite{bib:Paper2}, 
here we compensate the missing \chic\ polarization information by adopting the data-driven UAU conjecture:
direct \chic\ production is described by the same mixture of processes as the direct \jpsi\ and \psip\ production. 
In other words, the \pTovM\ shape functions of the polarized and unpolarized contributions, 
as well as the polarized fraction parameter $f_{\rm p}^{*}$, 
are assumed to be common to the P- and \mbox{S-wave} states.

As previously discussed, a possible realization of such a mechanism, 
not distinguishing between final states of different quantum numbers, 
is the production via colour-octet pre-resonances. 
This model inspires our definition of the possible \chic\ polarizations. 
In general, while an intrinsically unpolarized pre-resonance will lead to unpolarized S- and \mbox{P-wave} final states, 
a transversely polarized one will yield different polarizations when it transforms to 
\jpsi ($L = 0$, $J = 1$) or $\chi_{c0,c1,c2}$ ($L = 1$, $J = 0,1,2$) mesons, 
because of different changes in orbital and total angular momenta. 
The following expressions, only reflecting angular momentum conservation, 
relate the decay anisotropy parameter $\lambda_\vartheta$ of  \chic\ mesons 
(in the \jpsi\ plus photon channel) 
to the corresponding one of a \jpsi\ meson (in the dilepton decay channel), 
when all transform from the same pre-resonance state via gluon emissions:
\begin{align}
\label{eq:chiPolarization}
\begin{split}
\lambda_\vartheta^{\chi_0} &= 0 \; , \\
\lambda_\vartheta^{\chi_1} &= \lambda_\vartheta^{\jpsi} / (4 + \lambda_\vartheta^{\jpsi}) \; , \\
\lambda_\vartheta^{\chi_2} &= 21 \lambda_\vartheta^{\jpsi} / (60 + 13 \lambda_\vartheta^{\jpsi}) \; .
\end{split}
\end{align}
This means, for example, that the ``polarized'' \chic\ contribution, 
defined by $\lambda_\vartheta = +1$ for the \jpsi\ dilepton decay, 
is characterized by polarization parameters 0, $1/5$ and $21/73$, 
respectively, in the \chicZero, \chicOne\ and \chicTwo\ decays to \jpsi\ plus photon.

In our analysis we model the charmonium feed-down decay chains
($\psip \to \chi_{c1,c2} \gamma$,
$\psip \to \jpsi + X$ and 
$\chi_{c1,c2} \to \jpsi \gamma$)
taking into account momentum and polarization transformations.
Calculating the decay kinematics gives the relation 
connecting the mother's and daughter's laboratory momenta,
\begin{equation}
\label{eq:feeddownscaling}
\frac{ \langle p^2 \rangle / m^2 }{ P^2 / M^2 }
= 1 + \frac{ (M^2-m^2)^2 }{ 4 m^2 P^2 }
      + \frac{ (M^2-m^2)^2 }{ 4 m^2 M^2 }  \times ( 1 + \langle \cos^2\Theta \rangle ) \; ,
\end{equation}
where $M$ ($m$) and $P$ ($p$) are, respectively, 
the mass and laboratory momentum of the mother (daughter) particle,
and $\Theta$ is the emission angle of the daughter in the mother's rest frame. 
The average symbols denote an integration over $\cos\Theta$, making linear terms in this variable disappear. 
The term $\langle \cos^2\Theta \rangle$ can vary between $1/5$ and $2/5$ for physical polarizations of the mother particle. 
In the calculation we only made the approximation that $m_X \ll \sqrt{M^2+m^2}$, 
where $m_X$ is the total mass of the accompanying decay particles 
(either a photon or pions, depending on the decay channel). 
The deviation from unity in the right term of the equation 
is never larger than a few percent for all relevant feed-down cases and for $P$ not much smaller than $M$: 
the ratio of laboratory momentum over mass remains practically unchanged, on average, in the transition from mother to daughter. 
Since the two particles have, for $p \gg (M-m)$, almost indistinguishable directions in the laboratory, 
the relation is also valid vectorially and we can assume $p_{\rm T}/m = P_{\rm T}/M$.
In these conditions, we can model the kinematics of an indirectly produced particle using the \pTovM distribution of its mother particle.
Therefore, we account for the momentum conversion along any feed-down chain by simply considering all observables as functions of \pTovM.

The total prompt cross sections of the states affected by feed-down are obtained as
\begin{align}
\label{eq:PromptXsect}
\begin{split}
\sigma^{\chi_{c(1,2)}}(\pTovM) &= \sigma^{\chi_{c(1,2}}_{\rm dir}(\pTovM) \\
                                                 &+ B(\psip \to \chi_{c(1,2)} \gamma) \, \sigma^{\psip}(\pTovM) \; , \\
\sigma^{\jpsi}(\pTovM) &= \sigma^{\jpsi}_{\rm dir}(\pTovM) \\
&+ B(\chi_{c1} \to \jpsi \gamma) \, \sigma^{\chi_{c1}}(\pTovM) \\
&+ B(\chi_{c2} \to \jpsi \gamma) \, \sigma^{\chi_{c2}}(\pTovM) \\
&+ B(\psip \to \jpsi X) \, \sigma^{\psip}(\pTovM) \; ,
\end{split}
\end{align}
while $\sigma = \sigma_{\rm dir}$ in the \psip\ case.

We now address the relevant polarization transfer rules. 
Measurements of the $\psip \to \jpsi\ \pi \pi$ decay angular 
distribution~\cite{bib:Bai:1999mj, bib:Besson:1984ha, bib:Alexander:1998dq}
reveal that, in very good approximation, the \jpsi\ retains the angular momentum alignment of the mother \psip, 
being the $\pi \pi$ system dominated by its $J = 0$ component. 
It can be assumed, therefore, that in the \psip\ to \jpsi\ feed-down 
the polarization is transmitted without modification,
and that the shape parameters of the dilepton decay distribution (in particular, $\lambda_\vartheta$) do not change.
%%%
Concerning the \jpsi\ mesons produced by \chic feed-down decays, 
it has been shown~\cite{bib:chiPol}
that, while the angular momentum projection changes from mother to daughter, 
the $\chic \to \jpsi + \gamma$ decay and the \jpsi\ to leptons decay have surprisingly identical angular distribution shapes,
the latter one having 
the advantage of being unaffected by the non-zero orbital angular momentum components of the photon. 
Therefore, also in this case (but for different reasons) the angular decay parameter $\lambda_\vartheta$ remains the same 
(even if obviously referring to two different decay channels) in the \chic\ to \jpsi transition.
%%%
Finally, the \chic\ mesons produced by radiative decays of the \psip 
(as well as the \jpsi\ mesons from the subsequent \chic\ radiative decays) 
have the same polarization parameters as in Eq.~\ref{eq:chiPolarization}, 
after replacing $\lambda_\vartheta^{\jpsi}$ with $\lambda_\vartheta^{\psip}$.

To calculate the observable polarizations of the \jpsi\ and \chic\ states, 
combining their direct and indirect contributions (Eq.~\ref{eq:PromptXsect}), 
we use the general addition rule from Ref.~\cite{bib:Faccioli-PRD-FrameInv},
which, in the example of two processes contributing with fractions $f^{(1)}$ and $f^{(2)}$ of the cross section, reads as
\begin{equation}
\label{eq:polAddRule}
\lambda_\vartheta^{(1)} \oplus \lambda_\vartheta^{(2)} =
\frac{ f^{(1)} \lambda_\vartheta^{(1)} / (3 + \lambda_\vartheta^{(1)})  
      + f^{(2)} \lambda_\vartheta^{(2)} / (3 + \lambda_\vartheta^{(2)}) }
 { f^{(1)} / (3 + \lambda_\vartheta^{(1)}) 
+ f^{(2)} / (3 + \lambda_\vartheta^{(2)}) } \; .
\end{equation}
 
The fit has four state-independent (universal) free parameters: 
the shape parameters $\gamma$, $\beta_{\rm u}$ and $\beta_{\rm p}$ 
of the unpolarized and polarized cross sections and the fractional polarized contribution $f_{\rm p}^{*}$ 
(at the reference point $(p_{\rm T}/M)^{*} = 2$). 
The other parameters are the normalizations of the direct production cross sections of the four charmonia.
\begin{figure*}[t]
\centering
\includegraphics[width=0.39\linewidth]{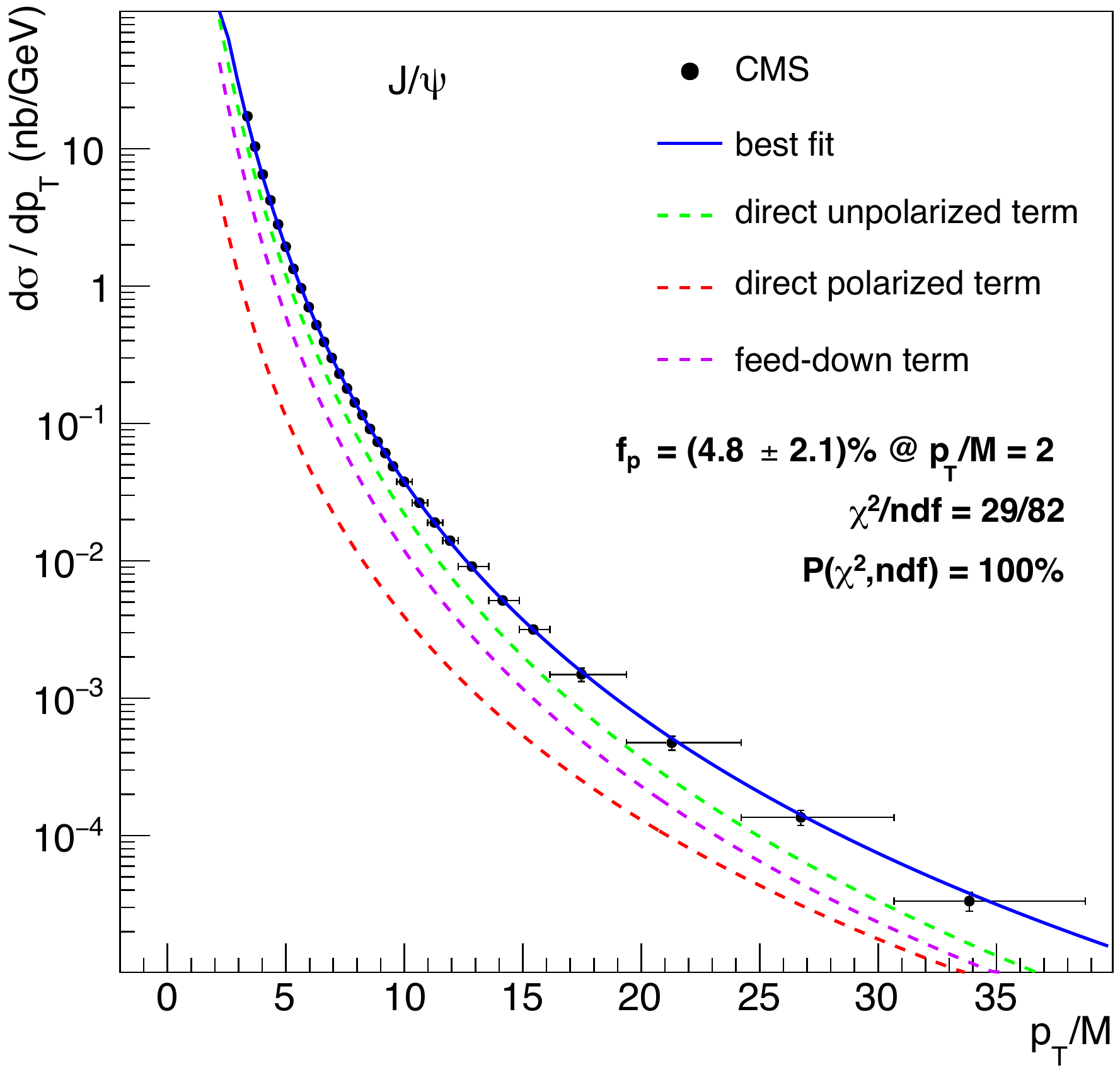}
\includegraphics[width=0.39\linewidth]{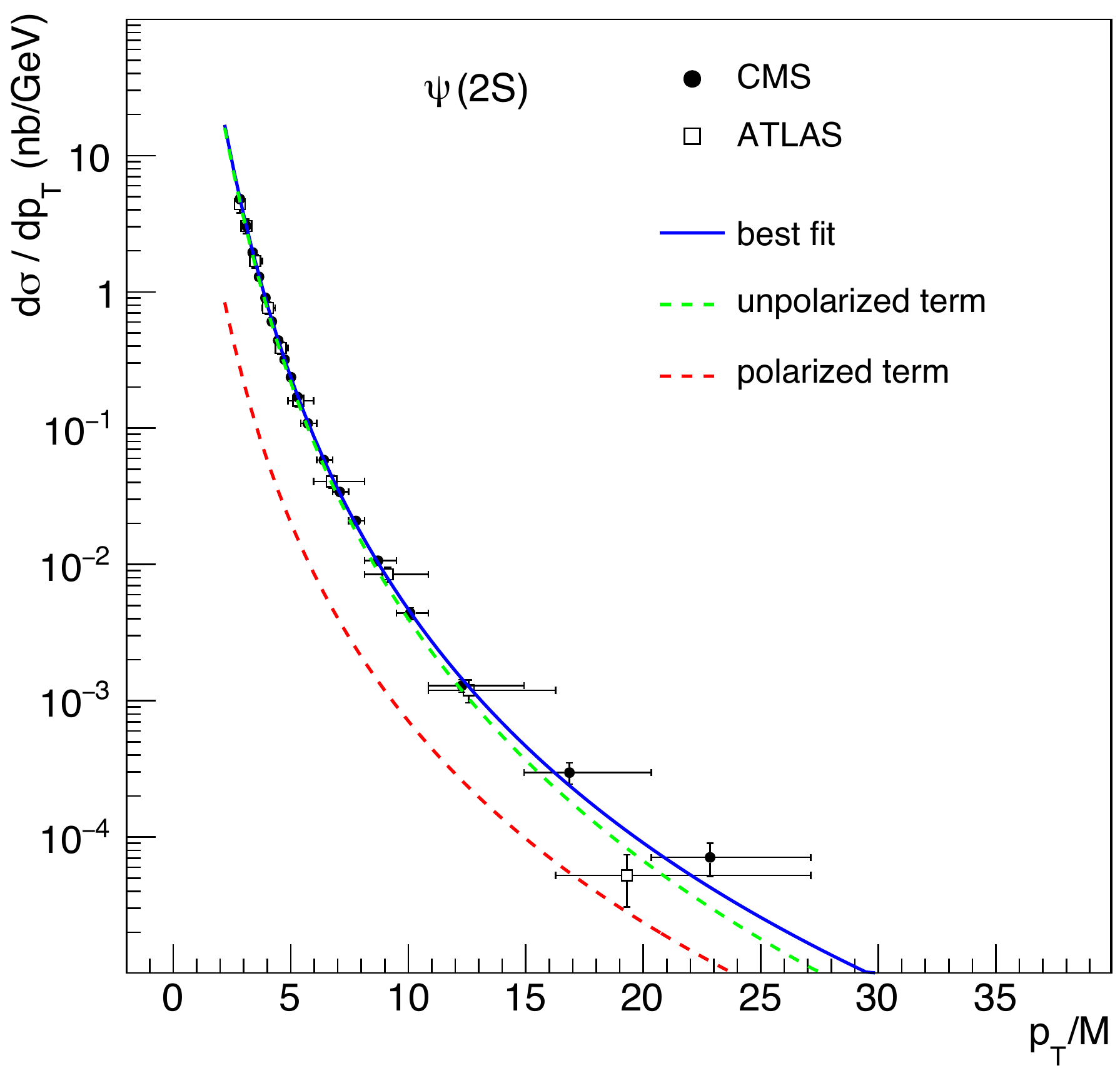}
\includegraphics[width=0.39\linewidth]{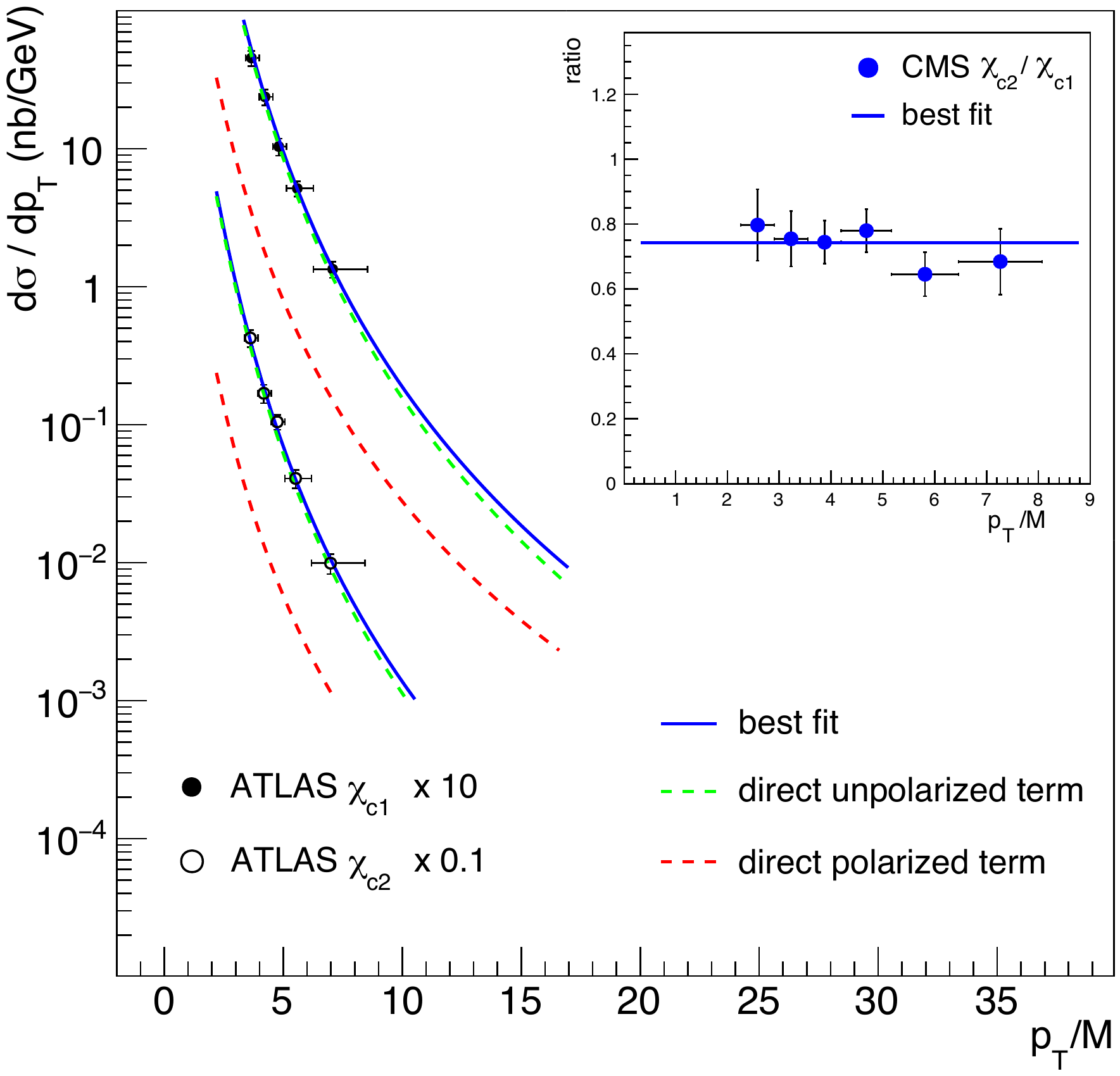}
\includegraphics[width=0.39\linewidth]{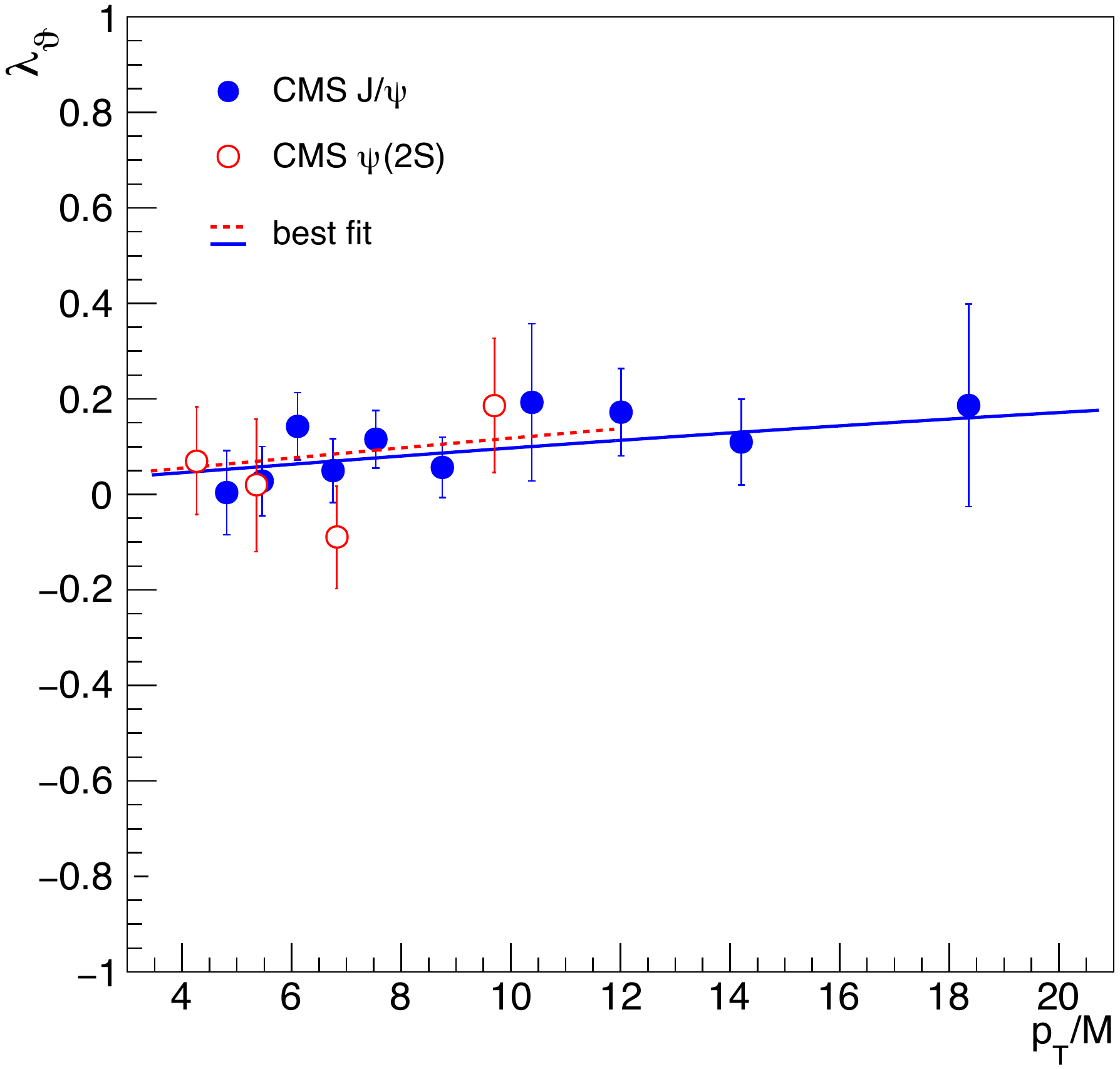}
\caption{Comparison between the data points and the best fit curves obtained in the UAU scenario:
\jpsi (top left) and \psip (top right) cross sections;
\chic\ cross sections and \chictwooverchicone ratio (bottom left);
and \lth\ for the \jpsi and \psip (bottom right).}
\label{fig:fittedDataUAU}
\end{figure*}

Additional nuisance parameters are used to model the correlations induced by the luminosity uncertainties. 
The branching ratios used to extract the cross sections from the yields measured in specific decay channels, 
as well as those weighting the corresponding feed-down terms of the prompt cross sections, 
are also treated as constrained parameters, with central values and uncertainties taken from Ref.~\cite{bib:PDG}.

The cross section measurements depend on the polarization hypothesis 
assumed by the experiments for the determination of the detection acceptance. 
For each set of parameter values scanned during the fit
and each \pt-rapidity bin of the cross section measurement,
we recalculate the average $\lambda_\vartheta$ parameter for the considered state and
apply the corresponding polarization-dependent acceptance correction.
The fit includes the charmonium measurements shown in Figs.~\ref{fig:pTovM}, \ref{fig:chiratios} and \ref{fig:polarizations}. 
The ATLAS \chic-to-\jpsi feed-down fraction and \chicTwo-to-\chicOne yield ratio are not considered, 
to avoid correlations with the \chic\ cross section data. 
The cross sections are only considered for $\pTovM > 2$.
The global fit has 82 degrees of freedom and leads to a total $\chi^2$ of 29, an exceptionally good result.
The agreement between the model and the data can be appreciated in Fig.~\ref{fig:fittedDataUAU}.

The polarized fraction is quite small at low \pTovM, 
reflecting the polarization data: $f_{\rm p}^{*} = (4.8 \pm 2.1)\%$ (at $(p_{\rm T}/M)^{*} = 2$).
It increases almost linearly with increasing \pTovM, becoming \mbox{$(20 \pm 5)\%$} at $\pTovM = 15$.
The best values of the shape parameters are $\gamma = 0.80 \pm 0.17$,
$\beta_{\rm u} = 3.47 \pm 0.05$ and $\beta_{\rm p} = 2.97 \pm 0.15$.

\begin{figure*}[t]
\centering
\includegraphics[width=0.8\linewidth]{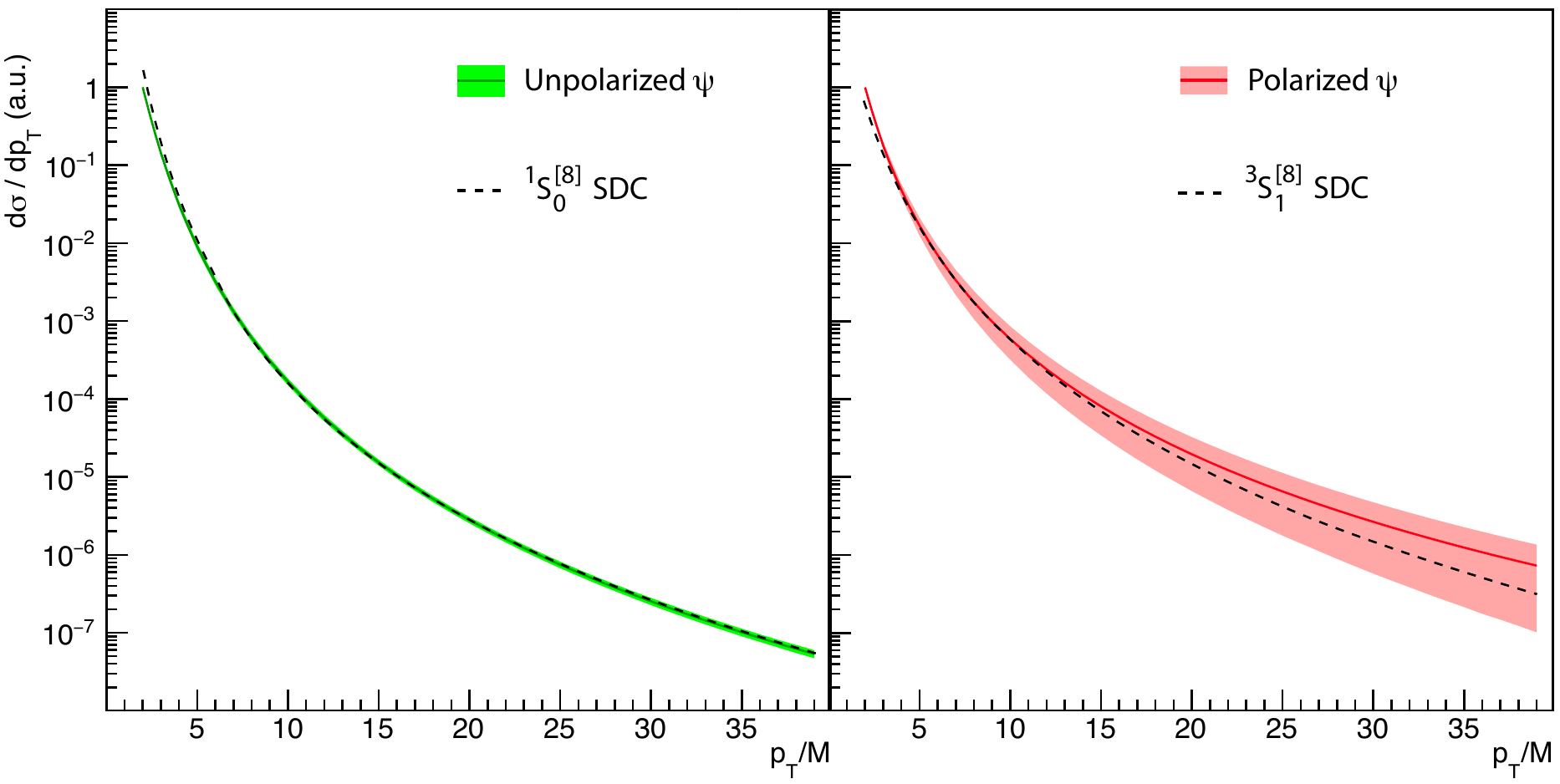}
\caption{Unpolarized (left) and polarized (right) cross sections in the UAU scenario, 
normalized to unity at $\pTovM=2$, with uncertainty bands reflecting correlated variations in the fit parameters.
The \oneSzero\ and \threeSone\ SDCs are also shown, arbitrarily normalized, 
calculated at NLO with NNLO fragmentation contributions~\cite{bib:BodwinCorrections}.}
\label{fig:pTovMdistrUAU}
\end{figure*}

Figure~\ref{fig:pTovMdistrUAU} shows the universal unpolarized and polarized \pTovM\ distributions obtained from the fit, 
normalized to unity at $\pTovM = 2$ to illustrate the allowed spectrum of kinematic shapes, irrespectively of the normalization shift.
The unpolarized curve (with a barely visible uncertainty band) is compared with the \oneSzero\ octet SDC shape 
calculated at NLO with NNLO fragmentation contributions, represented by the dashed curve: the agreement is astonishing.
Similarly, the \threeSone\ octet SDC calculation is in perfect agreement with the fitted polarized component,
affected by larger uncertainties.
This exercise supports the hypothesis that the universal unpolarized and polarized cross sections
have a physical affinity with the \oneSzero\ and \threeSone\ terms of the NRQCD factorization expansion.
It is worth noting that the two SDCs have significantly different shapes, 
their ratio changing by a factor 10 in the range $3 < \pTovM < 13$.

We conclude this section by discussing a new interesting hypothesis on
long-distance polarization effects in quarkonium production and
how it relates to the scenarios we are considering.
A crucial ingredient of NRQCD is the ``standard'' conjecture that the
\threeSone\ pre-resonance transfers its polarization (fully transverse at 
not-too-low \pt) to the observable \jpsi, \psip\ or \upsAllS\ without
attenuation, if no intermediate feed-down transitions are involved.
The process is modelled as a transition of the \QQbar\ pair with the
simultaneous emission of two
gluons having total angular momentum $J=0$, in analogy with the observable
process $\psip \to \jpsi\ \pi \pi$, where, as previously mentioned,
the \jpsi\ inherits the \psip\ polarization.
It has been recently observed~\cite{bib:Baranov:2016mka} that another kind
of process may be at play, consisting of two successive electric dipole transitions: 
$\threeSone \to \threePJ + g$, followed, for example, by $\threePJ\ \to \jpsi\ + g$. 
In this case, the resulting \jpsi\ and \upsAll\ polarization from \threeSone\
would be strongly attenuated with respect to the standard conjecture.
In fact, from the polarization perspective the two steps are
analogous to the radiative transitions $\psip \to \chi_c + g$ and
$\chi_c \to \jpsi\ + g$: in the first one the polarization is attenuated,
according to the relations in Eq.~\ref{eq:chiPolarization},
while the second step preserves the value of $\lambda_\vartheta$.
According to the calculations of Ref.~\cite{bib:Baranov:2016mka},
the \threeSone\ polarization may even be reduced to zero, if
interference effects are important. This ``double-transition'' process
should coexist with the one of the standard conjecture,
in presently unknown proportions, helping NRQCD predictions to converge to an unpolarized
scenario of \jpsi, \psip\ and \upsAllS\ production. The new hypothesis does not
affect \chic\ polarizations, since the transition of \threeSone\ to \chic\ is
already modelled according to the minimal assumption of a single electric
dipole transition.
%%%%%%%%%%%%%%%%%%%
A further possible justification of an unpolarized \threeSone
contribution has been provided in the $k_{\rm T}$-factorization approach~\cite{bib:Baranov2015psi2S}, 
which takes into account the off-shellness of the initial gluons.

Our general decomposition of \jpsi\ and \psip\ cross sections into ``unpolarized'' and (fully transversely) ``polarized'' contributions
can fit both scenarios, of transverse or unpolarized \threeSone,
since neither involves possible longitudinal contributions.
However, whenever a parallelism to the NRQCD \threeSone\ term is made,
as in Fig.~\ref{fig:pTovMdistrUAU}, we choose to refer to the standard
conjecture of \emph{fully} transverse \threeSone\ polarization.
In the new conjecture, instead, the \threeSone\ term would contribute 
to both the unpolarized ($\sigma_{\rm u}$) and polarized ($\sigma_{\rm p}$) cross sections,
depending on the effective polarization ($0 < \lambda_\vartheta(\threeSone) < 1$)
transmitted by the \threeSone octet to the \jpsi\ or \upsAllS\ meson.
The comparison in Fig.~\ref{fig:pTovMdistrUAU}, for example, should be made with 
a different association of the unpolarized and polarized contributions to the two octet channels,
\begin{align}
\label{eq:polUnpol3S1split}
\begin{split}
\sigma_{\rm u} & \rightarrow \sigma(\oneSzero) + \frac{4 - 4 \lambda_\vartheta(\threeSone)}{4 - \lambda_\vartheta(\threeSone)} \, \sigma(\threeSone) \; , \\
\sigma_{\rm p} & \rightarrow \frac{3 \lambda_\vartheta(\threeSone)}{4 - \lambda_\vartheta(\threeSone)} \, \sigma(\threeSone) \; ,
\end{split}
\end{align}
where $\sigma({^{\rm 2S+1}{\rm S}_J^{[8]}}) = 
\mathcal{S}({^{\rm 2S+1}{\rm S}_J^{[8]}}) 
\times \mathcal{L}({^{\rm 2S+1}{\rm S}_J^{[8]}})$.
The second equation shows that the polarized term should now be compared to a simply rescaled
\threeSone\ function, leaving the pure \emph{shape} comparison shown in the right panel unaltered. 
The comparison term for the unpolarized function, however,
would become a combination of the \oneSzero\ shape with a \threeSone\ contribution
depending on the assumed $\lambda_\vartheta(\threeSone)$ and \threeSone/\oneSzero yield ratios. 
Independently of these assumptions, such a combination would be a less steep function of \pTovM.
It is interesting to notice that, 
given the very good agreement between the unpolarized shape and the \oneSzero\ term alone 
(shown in Fig.~\ref{fig:pTovMdistrUAU}),
the present measurements already
disfavour a non-negligible contribution from an additional \emph{unpolarized} contribution 
having the shape of the \threeSone SDC.
In other words, the data do not indicate that the \threeSone polarization, 
always considered to be $\lambda_\vartheta(\threeSone) \approx +1$,
should instead be closer to~0.

%%%%%%%%%%%%%%%%%%%%%%%%%%%%%%%%%%%%
\section{Discussion and conclusions}
\label{sec:conclusions}

In this Letter we described a fully data-driven approach to the phenomenology of quarkonium production. 
We replaced perturbative calculations of the process kinematics, whose uncertainties are not reliably evaluated 
and in some cases are known to be large, with flexible empirical functions to be determined by the fit to the data. 
This choice assigns to the polarizations, mainly determined by basic physics considerations and, therefore, 
much more cleanly and stably predicted, 
the responsibility of discriminating between the different elementary processes of quarkonium production.

The proposed method implies an inversion of perspective in comparison to other analyses, 
where the imposition of the calculated individual cross-section contributions (the SDCs) 
results in a process discrimination based exclusively on the predicted \pt\ dependencies. 
In fact, polarization measurements are rarely included in global fits and, when included, 
they do not compete in constraining power with the cross section data, 
given the inferior statistical precision obtainable in multidimensional decay-angle analyses 
with respect to the much simpler yield counting. 
Our new fit procedure ignores the unobservable details of the theory calculations, 
in particular the existence of unphysical partial contributions (\mbox{P-wave} terms) 
and the factorization into the short- and long-distance matrix elements. 
The fit results exclusively reflect the data trends and uncertainties, remaining unbiased by theoretical ingredients.
Comparisons with SDC calculations and LDME evaluations are made a posteriori. 
This is a safer procedure for data-theory comparisons, 
separating the data-driven constraints, with their rigorous statistical interpretation, 
from the model calculations, usually affected by unreliable uncertainties.

The experimental picture of quarkonium production in pp collisions is still incomplete.
Once reasonably precise $\chi_c$ polarization measurements will become available,
our procedure will no longer need any model input.
At this moment, given the absence of $\chi_c$ polarization data,
we must resort to some model assumptions regarding $\chi_c$ production.
In this Letter, we assumed the UAU scenario, 
which simply reflects a full universality of charmonium production across all final states.

The disentangling of theory calculations from the fit to data 
allows us to face with awareness and devoted attention 
problems specific to one of the two heterogeneous worlds. 
For example, tensions originating from mutual experimental inconsistencies 
(such as those seen in Ref.~\cite{bib:ATLASdimuon})
become evident 
and can be addressed before and independently of the comparison to theory, 
without affecting the judgment of how well one or the other model describes the data.
Most importantly, 
while the fit results only represent the data with their well defined and statistically-interpretable uncertainties, 
they can be confronted \emph{simultaneously} to different models or multiple variants of the theory, 
the best way to appreciate the elusive concept of theory uncertainty. 
Instead, when model calculations are embedded in the fit, 
the quantification of theory uncertainties is either neglected or, in the best case, chosen rather arbitrarily 
(as in Ref.~\cite{bib:FaccioliPLB736}), 
possibly leading to subjective fit results, not amenable to rigorous interpretations.
Therefore, it is difficult to decide how well a model describes the data, 
since the goodness of the global fit 
(expressed by the $\chi^2$) 
has no statistical meaning when part of the uncertainties are lacking or are arbitrary.

In such cases, interesting indications can only come from macroscopic divergences,
seemingly beyond any reasonable uncertainty, as in the case of the famous polarization puzzle.
Even so, this puzzle persisted for so long exactly because of the 
entangled treatment of experimental and theoretical ingredients, 
which smeared the crucial (albeit a posteriori simple and predictable) evidence that 
the current perturbative SDC calculations, at NLO, are definitely steeper, at low \pt, 
than the corresponding experimental distributions. 
This feature, rigorously studied in Ref.~\cite{bib:FaccioliPLB736} 
by performing fits as a function of a selection cut progressively removing low-\pt data, 
is immediately visible when the results of our data-driven fits are compared to calculations, 
as done in Fig.~\ref{fig:pTovMdistrUAU}.
As also illustrated by the LDME determination presented in Ref.~\cite{bib:Paper2},
this method gives a better insight into the limitations of the current calculations, 
allowing us to quantify the effects of the implied approximations.

It is especially important to realize that the present experimental and theoretical knowledge 
on quarkonium production is significantly more satisfactory than in the past. 
In the absence of spectacular inconsistencies to be deciphered 
and with the prospect of forthcoming LHC measurements filling the few remaining gaps 
and further improving the constraining power of the experimental picture, 
quarkonium studies finally deserve the prestigious ranking of \emph{precision QCD studies}. 
Heading towards a detailed understanding of these phenomena, 
also through an accurate testing of NRQCD, a rigorous theory-data comparison method is essential.

Another central element in our analysis approach is the notion that 
particles characterized by the same partonic production mechanisms 
should have essentially the same $p/M$ distributions 
and that (for relatively small mass differences, as in the quarkonium spectrum) 
$p/M$ is also preserved in feed-down decays. 
This idea allows us to look for and identify physics patterns in a completely data-driven way.
For this reason, we have limited our present analysis to mid-rapidity LHC data, 
for which the scaling variable $p/M$ is sufficiently well approximated by \pTovM. 
The transformation of existing forward-rapidity \pt\ distributions into $p$ distributions 
requires a detailed analysis of rapidity dependencies and will be the subject of a future study.

The analysis reported in this Letter shows a surprising fact:
even the latest CMS and ATLAS cross-section measurements, 
extending up to \pt\ of order 100\,GeV, 
do not provide any evidence of differences in short-distance production mechanisms 
throughout the quarkonium spectrum. 
The \pTovM cross-section scaling is perfectly replicated among the seven measured states,
from the \jpsi to the \upsThree, 
and all five \mbox{S-wave} states, for which the polarizations have been measured, 
show similar unpolarized behaviours. 
While it is true that the $\chi$ states still require more precise and complete direct measurements 
(in particular, first polarization measurements), 
significant indications can be derived from the comparison of the kinematic distributions 
of the several \mbox{S-wave} particles, which receive very different contributions from $\chi$ decays. 
In fact, the measured \pTovM-differential \jpsi cross section, 
25\% of which arises from $\chi_c$ feed-down decays, 
is so similar to the \psip one, fully directly produced, 
that the $\chi_c$ \pTovM distribution is constrained to also have a very similar shape.
The significance of this observation is discussed in Ref.~\cite{bib:Paper2}, through a test where 
the \chicOne\ and \chicTwo\ spectra are replaced with single (average) points, 
thereby removing the experimental constraints on the $\chi_c$ cross section shapes,
which are then freely and independently determined by the global charmonium fit.

The idea that all measured quarkonia may have an undistinguishable partonic origin would imply that 
their final-state identities are acquired only in a subsequent, long-distance phase. 
This concept, a cornerstone of the NRQCD factorization framework, 
can be made extreme by assuming that all states, 
including the $\chi$ as much as the $\psi$ and $\Upsilon$ mesons, 
are predominantly produced from the same initial pre-resonance state, necessarily a colour-octet one. 
This is the hypothesis (UAU scenario) that we tested in this Letter, 
finding an excellent agreement with charmonium data. 
The existence of such a simple and satisfactory description of the data is quite remarkable and 
should stimulate theoretical 
considerations towards a natural explanation of the observed simplicity. 

Interestingly, the UAU conjecture, as an exclusive alternative to the 
NRQCD heavy-quark-limit symmetry rules, 
can be tested in lighter quarkonium systems. 
For example, the observation of almost unpolarized $\phi$ production,
with no significant dependence on the particle's laboratory momentum, 
would support the existence of fundamental principles dictating
that quark-antiquark bound-state formation mainly proceeds through unpolarized configurations.
The general principles implicit in the UAU hypothesis
would also naturally justify the seemingly surprising observation that the $\eta_c$ cross section 
only contains the ${^1{\rm S}_0^{[1]}}$ singlet term. 
When interpreted as the suppression of the ``higher-order'' members 
of an unexpectedly strong angular-momentum hierarchy, 
this observation becomes equivalent to the dominance of the ${^1{\rm S}_0^{[8]}}$
term in \jpsi production, another observation surpassing NRQCD expectations. 

Finally, by condensing all the experimental information on quarkonium production 
(at least at not-too-low \pt) in a simple analytical formula with few parameters,
the UAU scenario provides a useful empirical tool for Monte Carlo simulations, 
as well as for the modelling of background or reference processes, 
for example in studies of nuclear modifications of quarkonium production 
in proton-nucleus and nucleus-nucleus collisions.

%\clearpage
\section*{Acknowledgements}
We are indebted to Hua-Sheng Shao,
who kindly gave us the NLO calculations of the short distance coefficients.
The work of P.F.\ is supported by FCT, Portugal,
through the grant SFRH/BPD/98595/2013,
while the work of I.K.\ is supported by FWF, Austria,
through the grant  P\,28411-N36.

\section*{References}
\bibliographystyle{cl_unsrt}
\bibliography{GlobalFit2016}{}

\providecommand{\href}[2]{#2}\begingroup\raggedright\begin{thebibliography}{10}%
\makeatletter
\providecommand{\hrefCMSnoop }[0]{\@secondoftwo}%
\makeatother
\providecommand{\doi}{\texttt{doi:}\begingroup \urlstyle{tt}\Url}

\bibitem{bib:NRQCD}
\hrefCMSnoop {}{G.~T. Bodwin, E.~Braaten, and P.~Lepage} \textit{ Phys. Rev.}
  \textbf{ D51} (1995) 1125, \href{http://dx.doi.org/10.1103/PhysRevD.55.5853,
  10.1103/PhysRevD.51.1125}{\doi{10.1103/PhysRevD.55.5853,
  10.1103/PhysRevD.51.1125}},
  \href{http://www.arXiv.org/abs/hep-ph/9407339}{\texttt{arXiv:hep-ph/9407339}}.
[Erratum: Phys. Rev.D55,5853(1997)].
%%CITATION = HEP-PH/9407339;%%.

\bibitem{bib:FaccioliPLB736}
P.~Faccioli\hrefCMSnoop {}{ {et~al.}} \textit{ Phys. Lett.} \textbf{ B736}
  (2014) 98,
  \href{http://dx.doi.org/10.1016/j.physletb.2014.07.006}{\doi{10.1016/j.physletb.2014.07.006}},
\href{http://www.arXiv.org/abs/1403.3970}{\texttt{arXiv:1403.3970}}.
%%CITATION = ARXIV:1403.3970;%%.

\bibitem{bib:ATLASpsi2S}
\hrefCMSnoop {}{{ATLAS} Coll.} \textit{ JHEP} \textbf{ 09} (2014) 079,
  \href{http://dx.doi.org/10.1007/JHEP09(2014)079}{\doi{10.1007/JHEP09(2014)079}},
\href{http://www.arXiv.org/abs/1407.5532}{\texttt{arXiv:1407.5532}}.
%%CITATION = ARXIV:1407.5532;%%.

\bibitem{bib:ATLASYnS}
\hrefCMSnoop {}{{ATLAS} Coll.} \textit{ Phys. Rev.} \textbf{ D87} (2013)
  052004,
  \href{http://dx.doi.org/10.1103/PhysRevD.87.052004}{\doi{10.1103/PhysRevD.87.052004}},
\href{http://www.arXiv.org/abs/1211.7255}{\texttt{arXiv:1211.7255}}.
%%CITATION = ARXIV:1211.7255;%%.

\bibitem{bib:ATLASchic}
\hrefCMSnoop {}{{ATLAS} Coll.} \textit{ JHEP} \textbf{ 07} (2014) 154,
  \href{http://dx.doi.org/10.1007/JHEP07(2014)154}{\doi{10.1007/JHEP07(2014)154}},
\href{http://www.arXiv.org/abs/1404.7035}{\texttt{arXiv:1404.7035}}.
%%CITATION = ARXIV:1404.7035;%%.

\bibitem{bib:CMSjpsi}
\hrefCMSnoop {}{{CMS} Coll.} \textit{ Phys. Rev. Lett.} \textbf{ 114} (2015)
  191802,
  \href{http://dx.doi.org/10.1103/PhysRevLett.114.191802}{\doi{10.1103/PhysRevLett.114.191802}},
\href{http://www.arXiv.org/abs/1502.04155}{\texttt{arXiv:1502.04155}}.
%%CITATION = ARXIV:1502.04155;%%.

\bibitem{bib:CMSYnS}
\hrefCMSnoop {}{{CMS} Coll.} \textit{ Phys. Lett.} \textbf{ B749} (2015) 14,
  \href{http://dx.doi.org/10.1016/j.physletb.2015.07.037}{\doi{10.1016/j.physletb.2015.07.037}},
\href{http://www.arXiv.org/abs/1501.07750}{\texttt{arXiv:1501.07750}}.
%%CITATION = ARXIV:1501.07750;%%.

\bibitem{bib:Paper2}
P.~Faccioli\hrefCMSnoop {}{ {et~al.}} \textit{ Submitted to Phys. Lett.}
  \textbf{ B}
\href{http://www.arXiv.org/abs/XXX.YYY}{\texttt{arXiv:XXX.YYY}}.
%%CITATION = ARXIV:1403.3970;%%.

\bibitem{bib:CMSchic}
\hrefCMSnoop {}{{CMS} Coll.} \textit{ Eur. Phys. J.} \textbf{ C72} (2012) 2251,
  \href{http://dx.doi.org/10.1140/epjc/s10052-012-2251-3}{\doi{10.1140/epjc/s10052-012-2251-3}},
\href{http://www.arXiv.org/abs/1210.0875}{\texttt{arXiv:1210.0875}}.
%%CITATION = ARXIV:1210.0875;%%.

\bibitem{bib:CMSchib}
\hrefCMSnoop {}{{CMS} Coll.} \textit{ Phys. Lett.} \textbf{ B743} (2015) 383,
  \href{http://dx.doi.org/10.1016/j.physletb.2015.02.048}{\doi{10.1016/j.physletb.2015.02.048}},
\href{http://www.arXiv.org/abs/1409.5761}{\texttt{arXiv:1409.5761}}.
%%CITATION = ARXIV:1409.5761;%%.

\bibitem{bib:LHCbChibFeedown}
\hrefCMSnoop {}{{LHCb} Coll.} \textit{ Eur. Phys. J.} \textbf{ C74} (2014)
  3092,
  \href{http://dx.doi.org/10.1140/epjc/s10052-014-3092-z}{\doi{10.1140/epjc/s10052-014-3092-z}},
\href{http://www.arXiv.org/abs/1407.7734}{\texttt{arXiv:1407.7734}}.
%%CITATION = ARXIV:1407.7734;%%.

\bibitem{bib:LHCb_chicRatio}
\hrefCMSnoop {}{{LHCb} Coll.} \textit{ Phys. Lett.} \textbf{ B714} (2012) 215,
  \href{http://dx.doi.org/10.1016/j.physletb.2012.06.077}{\doi{10.1016/j.physletb.2012.06.077}},
\href{http://www.arXiv.org/abs/1202.1080}{\texttt{arXiv:1202.1080}}.
%%CITATION = ARXIV:1202.1080;%%.

\bibitem{bib:LHCb_chicJpsiRatio}
\hrefCMSnoop {}{{LHCb} Coll.} \textit{ Phys. Lett.} \textbf{ B718} (2012) 431,
  \href{http://dx.doi.org/10.1016/j.physletb.2012.10.068}{\doi{10.1016/j.physletb.2012.10.068}},
\href{http://www.arXiv.org/abs/1204.1462}{\texttt{arXiv:1204.1462}}.
%%CITATION = ARXIV:1204.1462;%%.

\bibitem{bib:LHCb_psip_pol}
\hrefCMSnoop {}{{LHCb} Coll.} \textit{ Eur. Phys. J.} \textbf{ C74} (2014)
  2872,
  \href{http://dx.doi.org/10.1140/epjc/s10052-014-2872-9}{\doi{10.1140/epjc/s10052-014-2872-9}},
\href{http://www.arXiv.org/abs/1403.1339}{\texttt{arXiv:1403.1339}}.
%%CITATION = ARXIV:1403.1339;%%.

\bibitem{bib:EPJC69}
\hrefCMSnoop {}{P.~Faccioli, C.~Louren\c{c}o, J.~Seixas, and H.~W{\"o}hri}
  \textit{ Eur. Phys. J.} \textbf{ C69} (2010) 657,
  \href{http://dx.doi.org/10.1140/epjc/s10052-010-1420-5}{\doi{10.1140/epjc/s10052-010-1420-5}},
\href{http://www.arXiv.org/abs/1006.2738}{\texttt{arXiv:1006.2738}}.
%%CITATION = ARXIV:1006.2738;%%.

\bibitem{bib:CMSlambdaPsi2S}
\hrefCMSnoop {}{{CMS} Coll.} \textit{ Phys. Lett.} \textbf{ B727} (2013) 381,
  \href{http://dx.doi.org/10.1016/j.physletb.2013.10.055}{\doi{10.1016/j.physletb.2013.10.055}},
\href{http://www.arXiv.org/abs/1307.6070}{\texttt{arXiv:1307.6070}}.
%%CITATION = ARXIV:1307.6070;%%.

\bibitem{bib:CMSlambdaYnS}
\hrefCMSnoop {}{{CMS} Coll.} \textit{ Phys. Rev. Lett.} \textbf{ 110} (2013)
  081802,
  \href{http://dx.doi.org/10.1103/PhysRevLett.110.081802}{\doi{10.1103/PhysRevLett.110.081802}},
\href{http://www.arXiv.org/abs/1209.2922}{\texttt{arXiv:1209.2922}}.
%%CITATION = ARXIV:1209.2922;%%.

\bibitem{bib:CDFpsisRunI}
\hrefCMSnoop {}{{CDF} Coll.} \textit{ Phys. Rev. Lett.} \textbf{ 79} (1997)
  572,
\href{http://dx.doi.org/10.1103/PhysRevLett.79.572}{\doi{10.1103/PhysRevLett.79.572}}.
%%CITATION = PRLTA,79,572;%%.

\bibitem{bib:etac_Kniehl}
\hrefCMSnoop {}{M.~Butensch{\"o}n, Z.-G. He, and B.~Kniehl} \textit{ Phys. Rev.
  Lett.} \textbf{ 114} (2015) 092004,
  \href{http://dx.doi.org/10.1103/PhysRevLett.114.092004}{\doi{10.1103/PhysRevLett.114.092004}},
\href{http://www.arXiv.org/abs/1411.5287}{\texttt{arXiv:1411.5287}}.
%%CITATION = ARXIV:1411.5287;%%.

\bibitem{bib:Chao:2012iv}
K.-T. Chao\hrefCMSnoop {}{ {et~al.}} \textit{ Phys. Rev. Lett.} \textbf{ 108}
  (2012) 242004,
  \href{http://dx.doi.org/10.1103/PhysRevLett.108.242004}{\doi{10.1103/PhysRevLett.108.242004}},
\href{http://www.arXiv.org/abs/1201.2675}{\texttt{arXiv:1201.2675}}.
%%CITATION = ARXIV:1201.2675;%%.

\bibitem{bib:Shao:2014fca}
\hrefCMSnoop {}{H.-S. Shao, Y.-Q. Ma, K.~Wang, and K.-T. Chao} \textit{ Phys.
  Rev. Lett.} \textbf{ 112} (2014) 182003,
  \href{http://dx.doi.org/10.1103/PhysRevLett.112.182003}{\doi{10.1103/PhysRevLett.112.182003}},
\href{http://www.arXiv.org/abs/1402.2913}{\texttt{arXiv:1402.2913}}.
%%CITATION = ARXIV:1402.2913;%%.

\bibitem{bib:Shao:2015vga}
\hrefCMSnoop {}{H.-S. Shao} \textit{ Comput. Phys. Commun.} \textbf{ 198}
  (2016) 238,
  \href{http://dx.doi.org/10.1016/j.cpc.2015.09.011}{\doi{10.1016/j.cpc.2015.09.011}},
\href{http://www.arXiv.org/abs/1507.03435}{\texttt{arXiv:1507.03435}}.
%%CITATION = ARXIV:1507.03435;%%.

\bibitem{bib:Petrelli}
A.~Petrelli\hrefCMSnoop {}{ {et~al.}} \textit{ Nucl. Phys.} \textbf{ B514}
  (1998) 245,
  \href{http://dx.doi.org/10.1016/S0550-3213(97)00801-8}{\doi{10.1016/S0550-3213(97)00801-8}},
\href{http://www.arXiv.org/abs/hep-ph/9707223}{\texttt{arXiv:hep-ph/9707223}}.
%%CITATION = HEP-PH/9707223;%%.

\bibitem{bib:BKNPB}
\hrefCMSnoop {}{M.~Butensch{\"o}n and B.~Kniehl} \textit{ Nucl. Phys. Proc.
  Suppl.} \textbf{ 222} (2012) 151,
  \href{http://dx.doi.org/10.1016/j.nuclphysbps.2012.03.016}{\doi{10.1016/j.nuclphysbps.2012.03.016}},
\href{http://www.arXiv.org/abs/1201.3862}{\texttt{arXiv:1201.3862}}.
%%CITATION = ARXIV:1201.3862;%%.

\bibitem{bib:BKMPLA}
\hrefCMSnoop {}{M.~Butensch{\"o}n and B.~Kniehl} \textit{ Mod. Phys. Lett.}
  \textbf{ A28} (2013) 1350027,
  \href{http://dx.doi.org/10.1142/S0217732313500272}{\doi{10.1142/S0217732313500272}},
\href{http://www.arXiv.org/abs/1212.2037}{\texttt{arXiv:1212.2037}}.
%%CITATION = ARXIV:1212.2037;%%.

\bibitem{bib:BKmodel}
\hrefCMSnoop {}{M.~Butensch{\"o}n and B.~Kniehl} \textit{ Phys. Rev. Lett.}
  \textbf{ 108} (2012) 172002,
  \href{http://dx.doi.org/10.1103/PhysRevLett.108.172002}{\doi{10.1103/PhysRevLett.108.172002}},
\href{http://www.arXiv.org/abs/1201.1872}{\texttt{arXiv:1201.1872}}.
%%CITATION = ARXIV:1201.1872;%%.

\bibitem{bib:BodwinCorrections}
\hrefCMSnoop {}{G.~T. Bodwin, H.~S. Chung, U.-R. Kim, and J.~Lee} \textit{
  Phys. Rev. Lett.} \textbf{ 113} (2014) 022001,
  \href{http://dx.doi.org/10.1103/PhysRevLett.113.022001}{\doi{10.1103/PhysRevLett.113.022001}},
\href{http://www.arXiv.org/abs/1403.3612}{\texttt{arXiv:1403.3612}}.
%%CITATION = ARXIV:1403.3612;%%.

\bibitem{bib:BaranovChi2011}
\hrefCMSnoop {}{S.~P. Baranov} \textit{ Phys. Rev.} \textbf{ D83} (2011)
  034035,
\href{http://dx.doi.org/10.1103/PhysRevD.83.034035}{\doi{10.1103/PhysRevD.83.034035}}.
%%CITATION = PHRVA,D83,034035;%%.

\bibitem{bib:BaranovChi2015}
\hrefCMSnoop {}{S.~P. Baranov, A.~V. Lipatov, and N.~P. Zotov} \textit{ Phys.
  Rev.} \textbf{ D93} (2016) 094012,
  \href{http://dx.doi.org/10.1103/PhysRevD.93.094012}{\doi{10.1103/PhysRevD.93.094012}},
\href{http://www.arXiv.org/abs/1510.02411}{\texttt{arXiv:1510.02411}}.
%%CITATION = ARXIV:1510.02411;%%.

\bibitem{bib:PDG}
\hrefCMSnoop {}{K.~A. Olive} \textit{ Chin. Phys.} \textbf{ C40} (2016) 100001,
\href{http://dx.doi.org/10.1088/1674-1137/40/10/100001}{\doi{10.1088/1674-1137/40/10/100001}}.
%%CITATION = CHPHD,C40,100001;%%.

\bibitem{bib:LHCb_etac}
\hrefCMSnoop {}{{LHCb} Coll.} \textit{ Eur. Phys. J.} \textbf{ C75} (2015) 311,
  \href{http://dx.doi.org/10.1140/epjc/s10052-015-3502-x}{\doi{10.1140/epjc/s10052-015-3502-x}},
\href{http://www.arXiv.org/abs/1409.3612}{\texttt{arXiv:1409.3612}}.
%%CITATION = ARXIV:1409.3612;%%.

\bibitem{bib:PhysRepHF}
\hrefCMSnoop {}{C.~Louren\c{c}o and H.~W{\"o}hri} \textit{ Phys. Rept.}
  \textbf{ 433} (2006) 127,
  \href{http://dx.doi.org/10.1016/j.physrep.2006.05.005}{\doi{10.1016/j.physrep.2006.05.005}},
\href{http://www.arXiv.org/abs/hep-ph/0609101}{\texttt{arXiv:hep-ph/0609101}}.
%%CITATION = HEP-PH/0609101;%%.

\bibitem{bib:pv}
\hrefCMSnoop {}{A.~David} \textit{ Phys. Lett.} \textbf{ B644} (2007) 224,
\href{http://dx.doi.org/10.1016/j.physletb.2006.10.071}{\doi{10.1016/j.physletb.2006.10.071}}.
%%CITATION = PHLTA,B644,224;%%.

\bibitem{bib:Bai:1999mj}
\hrefCMSnoop {}{{BES} Coll.} \textit{ Phys. Rev.} \textbf{ D62} (2000) 032002,
  \href{http://dx.doi.org/10.1103/PhysRevD.62.032002}{\doi{10.1103/PhysRevD.62.032002}},
\href{http://www.arXiv.org/abs/hep-ex/9909038}{\texttt{arXiv:hep-ex/9909038}}.
%%CITATION = HEP-EX/9909038;%%.

\bibitem{bib:Besson:1984ha}
\hrefCMSnoop {}{{CLEO} Coll.} \textit{ Phys. Rev.} \textbf{ D30} (1984) 1433,
\href{http://dx.doi.org/10.1103/PhysRevD.30.1433}{\doi{10.1103/PhysRevD.30.1433}}.
%%CITATION = PHRVA,D30,1433;%%.

\bibitem{bib:Alexander:1998dq}
\hrefCMSnoop {}{{CLEO} Coll.} \textit{ Phys. Rev.} \textbf{ D58} (1998) 052004,
  \href{http://dx.doi.org/10.1103/PhysRevD.58.052004}{\doi{10.1103/PhysRevD.58.052004}},
\href{http://www.arXiv.org/abs/hep-ex/9802024}{\texttt{arXiv:hep-ex/9802024}}.
%%CITATION = HEP-EX/9802024;%%.

\bibitem{bib:chiPol}
\hrefCMSnoop {}{P.~Faccioli, C.~Louren\c{c}o, J.~Seixas, and H.~W{\"o}hri}
  \textit{ Phys. Rev.} \textbf{ D83} (2011) 096001,
  \href{http://dx.doi.org/10.1103/PhysRevD.83.096001}{\doi{10.1103/PhysRevD.83.096001}},
\href{http://www.arXiv.org/abs/1103.4882}{\texttt{arXiv:1103.4882}}.
%%CITATION = ARXIV:1103.4882;%%.

\bibitem{bib:Faccioli-PRD-FrameInv}
\hrefCMSnoop {}{P.~Faccioli, C.~Louren\c{c}o, and J.~Seixas} \textit{ Phys.
  Rev.} \textbf{ D81} (2010) 111502,
  \href{http://dx.doi.org/10.1103/PhysRevD.81.111502}{\doi{10.1103/PhysRevD.81.111502}},
\href{http://www.arXiv.org/abs/1005.2855}{\texttt{arXiv:1005.2855}}.
%%CITATION = ARXIV:1005.2855;%%.

\bibitem{bib:Baranov:2016mka}
\hrefCMSnoop {}{S.~Baranov} \textit{ Phys. Rev.} \textbf{ D93} (2016) 054037,
\href{http://dx.doi.org/10.1103/PhysRevD.93.054037}{\doi{10.1103/PhysRevD.93.054037}}.
%%CITATION = PHRVA,D93,054037;%%.

\bibitem{bib:Baranov2015psi2S}
\hrefCMSnoop {}{S.~Baranov, A.~Lipatov, and N.~Zotov} \textit{ Eur. Phys. J.}
  \textbf{ C75} (2015) 455,
  \href{http://dx.doi.org/10.1140/epjc/s10052-015-3689-x}{\doi{10.1140/epjc/s10052-015-3689-x}},
\href{http://www.arXiv.org/abs/1508.05480}{\texttt{arXiv:1508.05480}}.
%%CITATION = ARXIV:1508.05480;%%.

\bibitem{bib:ATLASdimuon}
\hrefCMSnoop {}{{ATLAS} Coll.} \textit{ Eur. Phys. J.} \textbf{ C76} (2016)
  283,
  \href{http://dx.doi.org/10.1140/epjc/s10052-016-4050-8}{\doi{10.1140/epjc/s10052-016-4050-8}},
\href{http://www.arXiv.org/abs/1512.03657}{\texttt{arXiv:1512.03657}}.
%%CITATION = ARXIV:1512.03657;%%.

\end{thebibliography}\endgroup

\end{document}